\documentclass{article}

\usepackage{arxiv}

\usepackage[utf8]{inputenc} 
\usepackage[T1]{fontenc}    
\usepackage{hyperref}       
\usepackage{url}            
\usepackage{booktabs}       
\usepackage{amsfonts}       
\usepackage{nicefrac}       
\usepackage{microtype}      
\usepackage{lipsum}		
\usepackage{graphicx}
\usepackage{natbib}
\usepackage{doi}
\usepackage{mathptmx}
\usepackage{amsmath}
\mathchardef\mhyphen="2D
\usepackage{amssymb}
\usepackage[titletoc,toc,title]{appendix}
\usepackage{listings}
\usepackage{algorithm}
\usepackage{algpseudocode}
\newtheorem{definition}{Definition}[section]
\newtheorem{theorem}{Theorem}[section]
\newtheorem{lemma}{Lemma}[section]

\title{Sheaf-theoretic self-filtering network of low-cost sensors
for local air quality monitoring: A causal approach}

\author{{Anh-Duy Pham}\thanks{Corresponding author. Email: duyanhpham@outlook.com} \\
	Department of Computer Science\\
	Hochschule Bonn-Rhein-Sieg\\
	Sankt Augustin, Germany \\
	\And
	{Chuong Dinh Le} \\
	Department of Computer Science\\
	Vietnamese-German University\\
	Binh Duong, Vietnam \\
    \And
	{Hoang Viet Pham} \\
	Department of Computer Science\\
	Vietnamese-German University\\
	Binh Duong, Vietnam \\
    \And
	{Thinh Gia Tran} \\
	Department of Electrical and Computer Engineering\\
	Vietnamese-German University\\
	Binh Duong, Vietnam \\
    \And
	{Dat Thanh Vo} \\
	University of Windsor\\
	Windsor, Canada \\
    \And
	{Chau Long Tran} \\
	Aalto University\\
	Helsinki, Finland \\
    \And
	{An Dinh Le} \\
	Department of Electrical and Computer Engineering\\
	University of California San Diego\\
	California, USA\\
    \And
	{Hien Bich Vo} \\
	Department of Electrical and Computer Engineering\\
	Vietnamese-German University\\
	Binh Duong, Vietnam \\
}

\renewcommand{\headeright}{Article}
\renewcommand{\undertitle}{Article}
\renewcommand{\shorttitle}{Sheaf-theoretic self-filtering network of low-cost sensors}

\hypersetup{
pdftitle={Sheaf-theoretic self-filtering network of low-cost sensors for local air quality monitoring: A causal approach},
pdfsubject={cs.DC},
pdfauthor={Anh-Duy ~Pham, Chuong Dinh ~Le, Hoang Viet ~Pham, Thinh Gia ~Tran, Dat Thanh ~Vo, Chau Long ~Tran, An Dinh ~Le, Hien Bich ~Vo},
pdfkeywords={Sheaves, Consistency structure, Low-cost sensors, Air quality monitoring, Causality},
}

\begin{document}
\maketitle

\begin{abstract}
Sheaf theory, which is a complex but powerful tool supported by topological theory, offers more flexibility and precision than traditional graph theory when it comes to modeling relationships between multiple features. In the realm of air quality monitoring, this can be incredibly useful in detecting sudden changes in local dust particle density, which can be difficult to accurately measure using commercial instruments. Traditional methods for air quality measurement often rely on calibrating the measurement with public standard instruments or calculating the measurement’s moving average over a constant period. However, this can lead to an incorrect index at the measurement location, as well as an oversmoothing effect on the signal. In this study, we propose a compact device that uses sheaf theory to detect and count vehicles as a local air quality change-causing factor. By inferring the number of vehicles into the PM2.5 index and propagating it into the recorded PM2.5 index from low-cost air monitoring sensors such as PMS7003 and BME280, we can achieve self-correction in real-time. Plus, the sheaf-theoretic method allows for easy scaling to multiple nodes for further filtering effects. By implementing sheaf theory in air quality monitoring, we can overcome the limitations of traditional methods and provide more accurate and reliable results.
\end{abstract}


\section{Introduction}
The quality of the air we breathe is influenced by several variables, including the weather and traffic.
Measurements of air quality are done at many sites across the globe. This assists governments and communities in comprehending air pollution and implementing efforts to eliminate it. Many nations have monitoring stations for air pollution, and the majority have legislation to limit air pollution. Poor air quality causes several health issues, thus everyone should care about clean air. Globally, air quality monitoring is a frequent activity.
Numerous sensors are deployed in the air to monitor atmospheric contaminants. These measures aid in the comprehension of air pollution patterns and the development of remedies. The WHO SIDS (Sids) network and the EPA's National Air Quality Index are among the most widely used air quality assessment methods (AQI). SIDS takes data on temperature, relative humidity, and atmospheric pressure in order to compute an overall index of ambient air quality. The EPA index examines more than 500 distinct chemical components in the environment to evaluate if a region fulfills air quality guidelines. Each of these systems is distinct and monitors various variables, but they all have the same purpose: to give data for monitoring and the development of standards. 

To this extent, a representational framework with the ability to handle heterogeneous data is in increasing demand. There are numerous ways in which partial information about a particular occurrence might be conveyed. Monitoring of air quality is not an exception to this norm. For instance, the quantity of particles that can be inhaled (also known as the PM10 index), often having sizes of 10 micrometers or less; fine particles that can be inhaled (also known as the PM2.5 index), often having sizes of 2.5 micrometers or less; video footage from citizens; the number of vehicles on the road; or even Twitter texts could be used to measure and track air pollution. To reconstruct a complete image of what happens, it is necessary to blend these components in a way that is both understandable and effective. Because different observations have varying error models and may even be incompatible with one another, a model that is capable of integrating heterogeneous data forms has become critically necessary to satisfy sensor resource constraints while providing sufficient data to researchers. Fundamentally, we seek a thorough, theoretically motivated quantification of the agreement among sensor ensembles.

This study demonstrates, in conjunction with recent developments in signal processing \cite{robinson2014topological, robinson2017sheaves}, that mathematical sheaves allow an efficient processing framework with canonical analytic techniques. Due to the fact that just a minimal amount of hidden state must be estimated, relationships between sensors can be inferred by examining the consistency filtration, which is inherently established by the model. Sheaves are the archetypal structure of their kind, hence any systematic explanation of the interaction between diverse data sources will demonstrably recapitulate a portion of sheaf theory \cite{robinson2017sheaves}. Briefly, the theory gives an algorithmic method for globalizing data and diagnosing problems when data do not globalize cleanly. In addition, it demonstrates the feasibility of naturally introducing a supplementary element, such as vehicle numbers, as a causally directed treatment to the core signals for self-consistent adjustment of likely local signal peaks. This allows one to discover groups of self-consistent sensors and determine when or where particular sensor combinations are likely to be in agreement. We base our discussion of methodology on a real-world experiment. This experiment demonstrates that although the sheaf methodology requires rigorous modeling as a prerequisite, one may achieve fine-grained analytics that are automatically tuned to the particular sensor deployment and the set of observations. The experimental design and sheaf modeling are based on the convention described in Joslyn et al. \cite{joslyn2020sheaf}. This research addresses the multi-sensor fusion problem as opposed to a conventional causal inference since the causal aspect is implicitly baked into the sheaf network through the self-consistent process from the vehicle amount derived from one node to an air quality index node. In the framework of the research, we developed an air monitoring system and a specific vehicle tracking and counting system, therefore in addition to data fusion approaches, this section also discusses related work regarding the aforementioned methodologies. 

\subsection{Air monitoring systems}
Regarding the hardware system, a study in Thailand employs a similar methodology \cite{sahanavin2018relationship}: their primary objective was to record PM2.5, PM10, temperature, and humidity in addition to vehicle flow, which was categorized into five groups: light-duty gasoline vehicles, light-duty liquid petroleum gas (LPG) vehicles, heavy-duty diesel vehicles, and light-duty diesel vehicles. To collect particle data, a PTFE filter and an air pump were utilized, and the obtained matter was weighed on an ultra-microbalance with a minimum readability of $0.1 \mu g$. Temperature and humidity were recorded by a separate meteorological device. The traffic flow was finally captured by a closed-circuit camera. The association between traffic flow and particulate matter is a modest effect in indoor air quality, while temperature and relative humidity appear to have a significant impact. However, they noted that the restricted monitoring region could result in incorrect results. Moreover, Zhang et al. \cite{zhang2021low} in the United States created a system for indoor air quality monitors using Raspberry Pi as the central controller. They created a logging system for SO2, NO2, TVOC, O3, CO, CO2, PM2.5, PM10, temperature, and humidity. Using a USB connector, they also simplified the sensor and Raspberry Pi connection. Their Low-Cost Air Quality Sensors system, however, is designed for inside monitoring and cannot thus withstand outdoor circumstances. We-count proposes to measure mobility and air quality data using Raspberry Pi in Europe \cite{momirski2022southern}. The pilot tests have been launched in Madrid, Ljubljana, Dublin, Cardiff, and Leuven. From these nodes, the project can give a data set of vehicle and air quality data for scientists to study the association between traffic and air pollution. Two gadgets comprise the We-count project: a Raspberry Pi camera system and an IoT device for measuring PM. Volunteers install the visible data logging device on an interior glass window, while the dust data logging apparatus is installed outdoors beneath the window's roof. Two separate devices may complicate the mapping network.

\subsection{Vehicle to PM2.5 concentration}
Using both linear regression and route analysis, a 2018 study from Bangkok, Thailand \cite{sahanavin2018relationship} proposed investigating the association between PM2.5 level and traffic flow on numerous roadsides using both linear regression and path analysis. The outcome of statistical model study suggests that traffic flow has a direct impact on particle levels in both open and enclosed regions. The study also shown that route analysis provides more precise and efficient results than linear regression at both low and high PM concentration levels. A guideline from the European Environment Agency \cite{eeaeuropa} also covered the exhaust emissions, including PM2.5, N2O, CO2, NH3, etc., from road traffic depending on a variety of vehicle-related parameters. Together, the total distance driven by the cars, the technology-specific equipment used to measure the emission parameters, and the number of vehicles on the road are used to more precisely estimate the amount of related emissions. In addition, several tables including the estimated emission factor values for each commercial vehicle technology in Europe were produced by the study. Specifically for Ho Chi Minh city, Vietnam, there is a formula for calculating the amount of PM2.5 a vehicle emits, and it is as shown in \cite{phung2020development} thanks to the guidebook from a study conducted by the California Air Resource Board in 2020 \cite{californiaairboard}, which outlined the relationship between various vehicle model technologies and their pollutant emission, such as NOx, PM2.5, CO, and ROG. 

\subsection{Multi-sensor fusion and sheaf-theoretical approach}
To this extent, data fusion is the process of creating a "union of data from several sources" \cite{wald1999some}. Also reported Leal et al. and Solera et al. \cite{leal2015motchallenge, solera2015towards} are a variety of methodical experimental programs. Other authors have demonstrated that combining sensor detections \cite{bailer2014superior, hall2004mathematical, smith2006approaches, newman2013upstream} improves coverage and performance. Using a heterogeneous collection of sensors, some of which are manually operated, is analogous to the sensor deployment strategy described in \cite{zhu2017integrating}, which outlines a method for monitoring vehicular traffic on a road network using temporary, portable sensors. Although their fundamental model is linear and satisfies a conservation law, ours is not required to be linear. Most of the time for data fusion methods, a lot of work goes into making features that are strong, but the way features are chosen is usually based on their applicability rather than their theoretical value. In every case, though, it is assumed that the sample rates are high enough. Also, sensors of the same type are usually needed for data fusion techniques that work with quantitative data \cite{alparone2008multispectral, varshney1997multisensor, zhang2010multi}. Most of the time, sensors need to be registered to a common coordinate system before they can be used \cite{dawn2010remote, guo2008potential, koetz2007fusion}. 

Due to the absence of a standard coordinate system, all of these techniques are susceptible to changes in sensor placement. Using a more foundational method, such as so-called "possibilistic" information theory \cite{benferhat2006reasoning, benferhat2009fusion, crowley1993principles}, is one way to mitigate this difficulty. Here, sensor models are encoded as a collection of propositions and rules of inference. The data then determine the values of logical variables, from which conclusions can be derived about the situation. Possibilistically valuing variables, as opposed to probabilistically, enables a broader range of uncertainty models. Although these methods can typically accommodate heterogeneous collections of sensors, they do so without the theoretical guarantees one would anticipate from homogeneous collections. While the methodology for possibilistic techniques is comparable to ours, the fundamental difference is that we rely on geometry \cite{joslyn2020sheaf}. In the absence of a geometric structure, the combinatorial complexity rapidly increases with the number of possible sensor outputs, preventing the direct use of logical approaches.

Typical target tracking and data fusion strategies tend to postpone modeling until after observations are available. If this approach were reversed, so that careful modeling was required before any observations were examined, then sample requirements would be less stringent. This is backed by the proposed methodology, which employs local models of consistency among the data. Sheaves provide a canonical and practical formalization of these interconnecting local models. A sheaf is a detailed specification of which local data sets can be merged into a more consistent, global datum \cite{malcolm2009sheaves}. Indeed, sheaves are an efficient organization tool for deployments of heterogeneous sensors \cite{joslyn2014towards}. The most fundamental data fusion question addressed by a sheaf is whether a whole set of observations represents a global section, which is a perfectly consistent and unified state \cite{purvine2018topological}. The consistency radius \cite{robinson2017sheaves} quantifies the relationship between the geometry of a sheaf and a collection of observations. Practical algorithms for data fusion emerge naturally, for instance, as minimization methods for the consistency radius \cite{robinson2018dynamic}. As in the more theoretical treatment \cite{robinson2018assignments}, we demonstrate in this paper that sheaves allow further, finer-grained analysis of consistency via consistency filtration. Cohomology is a technical technique for understanding, for sufficiently structured sheaf models, how local observations can or cannot be fused. This has applications in numerous fields, including network structure \cite{robinson2016imaging, ghrist2011network, nguemo2017sheaf} and quantum information \cite{abramsky2015contextuality, abramsky2011sheaf}. Cohomology computation is straightforward \cite{robinson2014topological} and efficient techniques are known \cite{curry2016discrete}.

\section{System overview}
\label{section:network}
Overall, the system is composed of two key components: the air pollution sensing hardware and the PM2.5 aggregating algorithm. The hardware utilizes a range of low-cost air sensors and cameras to collect data, which is then processed by the algorithm to infer PM2.5 concentration levels. The algorithm subsection is further divided into two parts: the vehicle counting algorithm and the sheaf modeling approach. The vehicle counting algorithm uses advanced algorithms to accurately detect and count vehicles, which are then used as a factor in determining the PM2.5 concentration levels. Sheaf modeling is a complex but powerful tool that allows for more flexibility and precision in modeling relationships between multiple features. In the context of air quality monitoring, it can be used to more accurately detect sudden changes in local dust particle density. This approach is discussed in more detail in a separate section.

\subsection{Cost-effective air pollution sensing hardware}
The system is designed to gather a wide range of data, including traffic video footage, temperature, relative humidity, atmospheric pressure, and particulate matter (PM) levels. This comprehensive dataset will provide valuable insights into the factors that affect local air quality. The 3D-printed casing not only protects the system and its components, but also allows for easy deployment and maintenance. By deploying the system to roadside houses, we can gather data from a variety of locations, providing a more accurate and comprehensive view of air quality in a given area.

The electrical system at the heart of our system is powered by a Raspberry Pi 3B, which is equipped with a Pi Camera V2 module. This allows for the capture of high-resolution video data, providing valuable insights into local air quality. The Raspberry Pi is also connected to an Environmental Monitoring Module, which includes sensors to collect meteorological metrics such as temperature, relative humidity, and atmospheric pressure. This comprehensive dataset allows for a thorough understanding of local air quality conditions. To store and manage the large amounts of data generated by the system, we have chosen to use Azure Blob Storage as our cloud storage solution. This ensures the security and accessibility of the data, allowing for easy analysis and interpretation.

\begin{figure}[t!]
  \includegraphics[width=\linewidth]{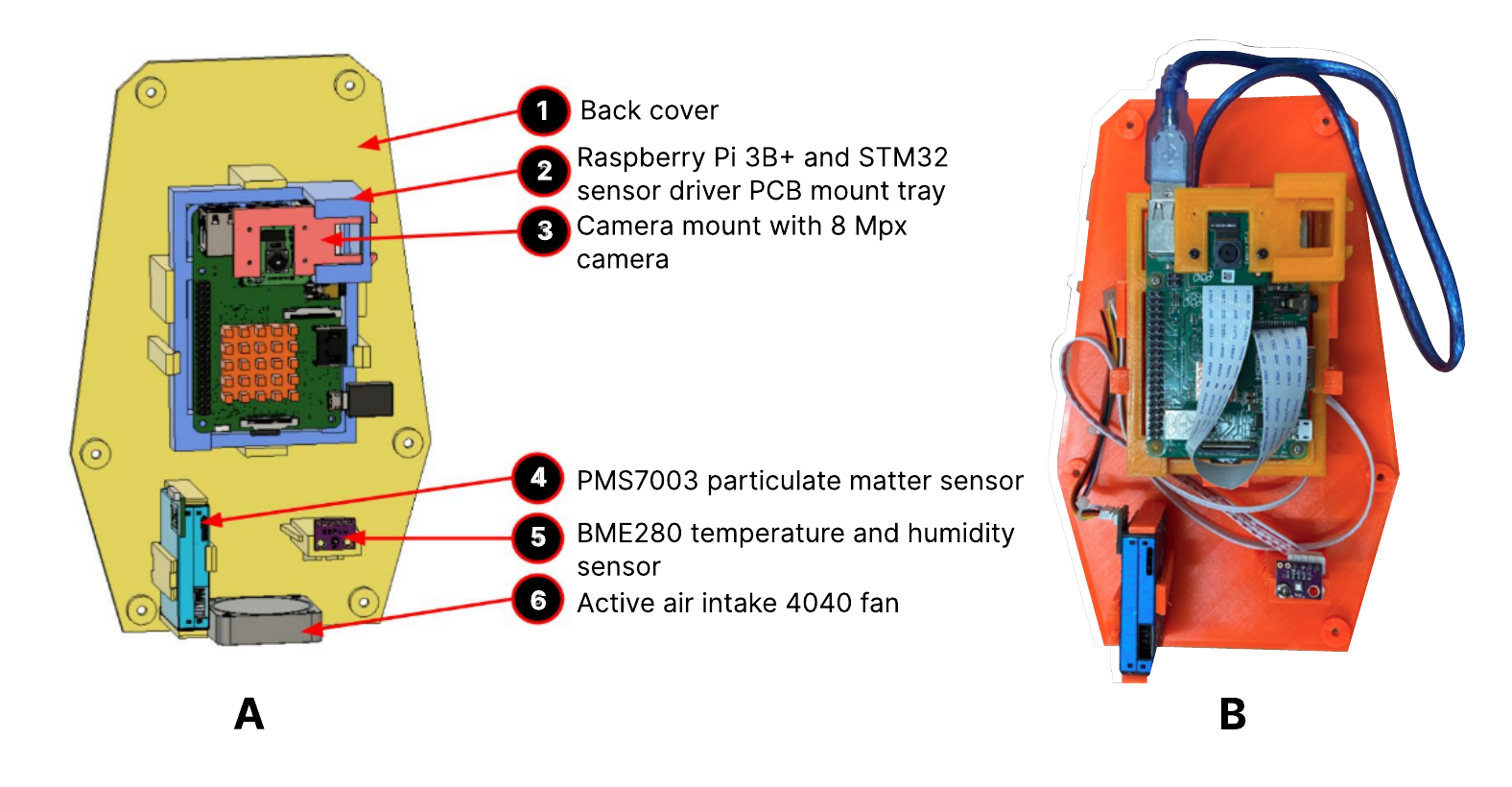}
  \caption{A single integrated air sensing hardware system. A - Computer-aided design, B - Real wired design.}
  \label{fig:hardwaresystem}
\end{figure}

Our system, allows for automated operation through a Python script, which captures high-quality video through the Pi Camera V2 and uploads the videos to Azure Blob Storage for further processing. It also collects data from the Environmental Monitoring Module and saves it to a CSV file, which is uploaded to Azure Blob Storage for storage and analysis. This automated approach ensures efficient and reliable data gathering, providing a comprehensive view of local air quality conditions. The use of the Raspberry Pi, Python script, and Azure Blob Storage enables a robust and scalable system for air quality monitoring. Figure \ref{fig:hardwaresystem} shows the overview of our system both in computer-aided design and real design. Specifically, the Environmental Monitoring Module consists of an STM32 Blue Pill development board with a range of sensors connected to it, including a DS1307 real-time clock circuit for real-time tracking, a BME-280 sensor for recording temperature, humidity, and atmospheric pressure data, and a PMS-7003 low-cost PM sensor for measuring PM1.0, PM2.5, and PM10 concentrations. The module also includes a Pi Camera V2 for recording video data of traffic at a resolution of 1296x730 pixels and 10 frames per second. The low-cost camera can be easily integrated into a Raspberry Pi for remote control. The data generated by the module will be stored in Azure Blob Storage, a low-cost and high-volume cloud storage solution from Microsoft. The Raspberry Pi will automatically divide the video files into 5-minute segments and upload them to Azure Blob Storage using the Azure API. It is estimated that a single node will generate around 30GB of data per day.

The core concept of our system is based on the work of Tran et al. \cite{https://doi.org/10.48550/arxiv.2212.04313}, but with an expanded scale to take advantage of the power of sheaf modeling. Instead of using a single, integrated system as described in their work, which only allows for two vertices in the sheaf model, we have expanded the model to include four main vertices: two dust sensors and two cameras. This dual system enables a more comprehensive and accurate analysis of local air quality conditions. Our sheaf model, discussed in more detail in Section \ref{sec:sheafmodel}, allows for the aggregation of data from multiple sources to provide a more complete view of the local air quality. It goes beyond simply yielding the statistical distribution of the two vertices, providing a more accurate and nuanced understanding of the factors affecting local air quality. 

\subsection{Vehicle counting and PM2.5 concentration conversion algorithms}
In order to determine the relationship between PM2.5 concentration and traffic density, it is necessary to count vehicles. To count the number of vehicles, we use a virtual line method. This involves placing a static line on the video frame, and then counting the number of vehicles that pass through it. However, this method has some limitations, as the camera's low sampling rate (24 fps) means that some vehicles may be missed. To address this, we have developed an additional algorithm that temporarily draws a rectangle around a vehicle once it has been counted, so that it will not be counted again. The rectangle disappears after a certain number of frames without detecting any vehicles. To improve the accuracy and reliability of our vehicle counting system, we have developed a number of additional algorithms and techniques. One of these is the use of the DenseNet-121 \cite{huang2018kilian} model for classification, which offers a good balance of accuracy and real-time performance. The model is trained on a dataset of around 1000 images of cars and motorcycles, and is able to accurately distinguish between the two classes.

In addition to these algorithms, we have also implemented a number of post-processing steps to improve the accuracy of our system. This includes resampling the data using the STL method with a 1 hour frequency to smooth out the curve, and applying a series of morphological transformations to the output of the background subtraction algorithm. This includes erosion and dilation using a 5x5 kernel filled with 1s, followed by a closing operation using a 5x5 ellipse-shaped kernel. This sequence of operations has been found to produce good results when contouring the output image.

\begin{figure}[t!]
  \centering
  \includegraphics[width=0.75\linewidth]{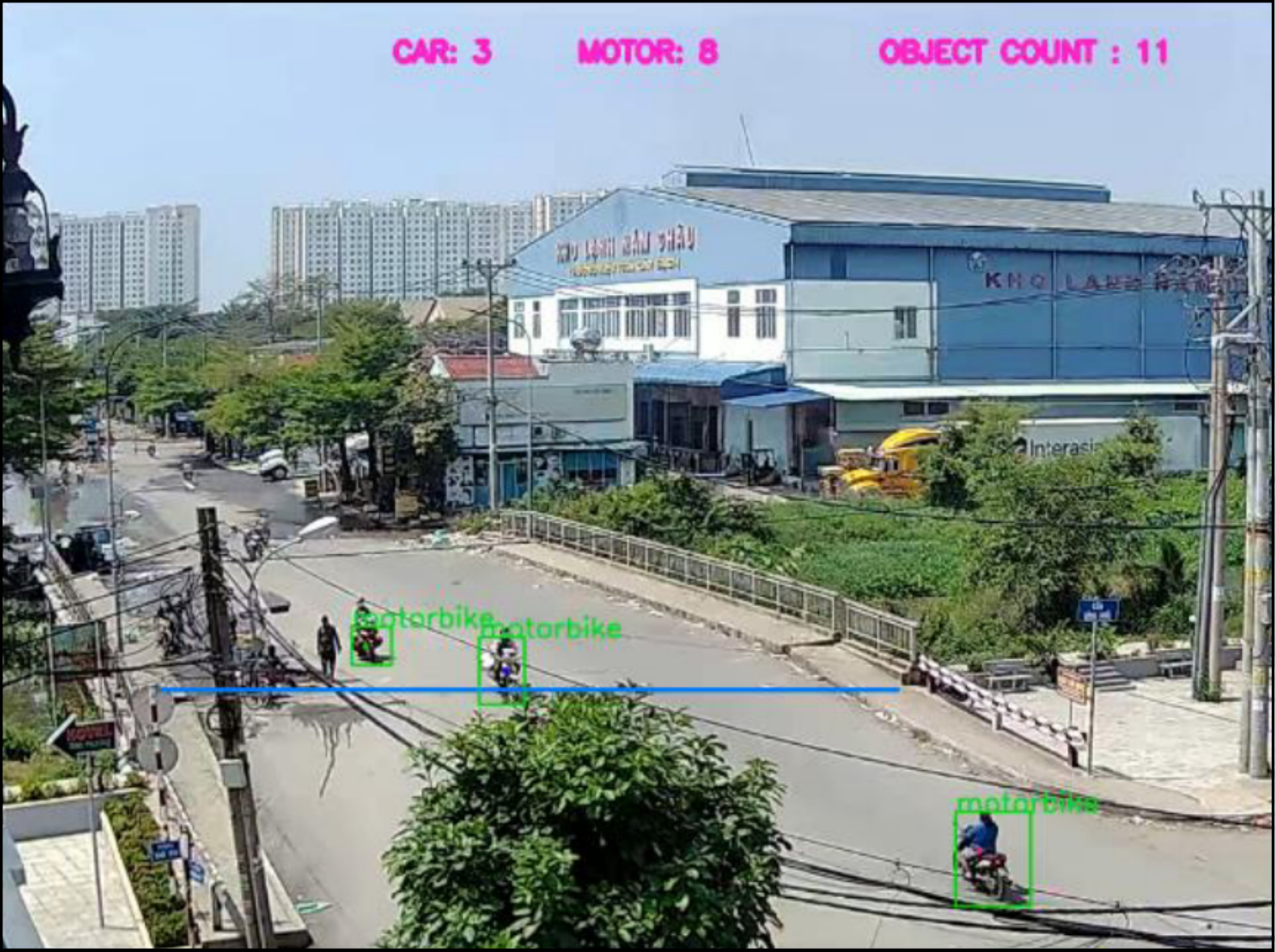}
  \caption{Demonstration of our tracking system in low-cost camera.}
  \label{fig:roadimg}
\end{figure}

Overall, our vehicle counting algorithm, which is adapted from the work of Le at al \cite{https://doi.org/10.48550/arxiv.2212.01761}, is able to accurately and efficiently count vehicles in real-time, providing valuable data for studying the relationship between PM2.5 concentration and traffic density. The in-field demonstration is shown in Fig. \ref{fig:roadimg}.

\begin{table}\centering
\caption{Emission factors of typical vehicles in Vietnam}
\label{tab:EF}
\begin{tabular}[t]{lc}
\hline \hline
            & Emission Factor $(g*km^{-1})$  \\
\hline
Two-wheeled vehicle  &   0.047  \\  
Four-wheeled vehicle &   0.117     \\
\hline
\end{tabular}
\end{table}

There is a formula for calculating the amount of PM2.5 emitted by vehicles, which is stated as follows according to Phung, Nguyen Ky, et al. \cite{phung2020development}:

\begin{equation}
\label{eq:Em}
E_m = N_m \times EF_m \times VKT_m
\end{equation}
, where $E_m$ is the mass of emitted PM2.5 $(g)$, $N_m$ is the number of vehicles type $m$, $EF_m$ is the emission factor of vehicle type $m$ $(g*km^{-1})$ and $VKT_m$ is the length of the recorded street segment $(km)$. The emission factor used in our calculations can be found in the California Air Resources Board Emission Factor Table \cite{californiaairboard}, which is summarized into two typical types of vehicle in Vietnam for our implementation in Table \ref{tab:EF}. Using this factor and formula \ref{eq:Em}, we can calculate the PM2.5 concentration by dividing the emission rate by the volume of air in which the particles are distributed. In our calculations, we assume that the PM2.5 particles are uniformly distributed in a cube with a side length equal to the vehicle kilometers traveled (VKT). This allows us to accurately estimate the PM2.5 concentration in a given area and time period as follows:

\begin{equation}
\label{eq:PM}
C_m = E_m \times 10^6 / (VKT_m \times 1000)^3
\end{equation}
, where $C_m$ is the target PM2.5 concentration $(\mu g/m^3)$.

We now have a formula to map the number of vehicles to PM2.5 concentration; however, it is clear that PM2.5 from vehicles cannot be detected by a sensor right away, but rather after a delay. To determine this delay, Le et al. \cite{https://doi.org/10.48550/arxiv.2212.01761} have developed a method that calculates the average delay of all local delays, and shift the converted PM2.5 concentration signal inferred from the number of vehicles based on this delay. We lift their approach one level further to better capture the dynamic of the two types of signal by determining the lag by cross-correlation method. Specifically, after each 24-hour period, the vehicle signal and the PM2.5 concentration signal from an air pollution sensor are associated with each other using cross-correlation to find the most viable lag among the data points. The lag with maximal correlation is then chosen for shifting the converted signal in the next day, which summarized in Algorithm \ref{alg:basePMlag}. However, vehicles are a source of PM2.5, but not the only one. In order to calculate the total PM2.5 concentration, we need to estimate the contribution of other sources. We assume that for a given hour, the contributions of other sources are constant. We take one day, calculate the pattern for that day, which is equalled to the subtraction of PM2.5 concentration read from the sensor and the lag-shifted PM2.5 conversion from the camera data, and use that pattern for the following day. To derive the total PM2.5 concentration of the next day, we add the pattern with the PM2.5 concentration from vehicles. The following algorithm concludes the overall process of converting an image containing vehicles to total PM2.5 concentration.

\begin{algorithm}
\caption{VehicleImage2PM2.5 (VI2PM) algorithm}\label{alg:vehicle2PM}
\begin{algorithmic}
\Require $I$: an image of a road containing vehicles, \\$\quad \quad \quad$ \textbf{$T_v$}: total vehicle types, \\ $\quad \quad \quad \ L_{t-1}$: lag factor between sensing and converted PM2.5 concentration of the previous day,  \\ $\quad \quad \quad \ P_{b,t-1,h}$: base PM2.5 concentration of other emission sources in the previous day $(t-1)$ of an hour $h$
\Ensure $P_{v,t,h}$: PM2.5 concentration of emission from vehicles in the current day $t$ of an hour $h$ \\ $\quad \quad \quad \ P_{T,t,h}$: aligned total PM2.5 concentration from the number of vehicles in the image in the current day $t$ of an hour $h$
\State $\{\textbf{O}\} \gets  \textbf{ModifiedBackgroundSubtraction}(I)$
\State $\{\textbf{M}_i: \textbf{O}_j\} \gets \textbf{DenseNet}(O_j)$ \textbf{for} $i$ \textbf{in} \textbf{$T_v$}
\State $\{N_m\} \gets \textbf{LineCounting}(m)$ \textbf{for} $m$ \textbf{in} \textbf{$M$}
\State $\{E_m\} \gets N_m \times EF_m \times VKT_m$ \textbf{for} $m$ \textbf{in} \textbf{$M$}
\State $\{C_m\} \gets E_m \times 10^6 / (VKT_m \times 1000)^3$ \textbf{for} $m$ \textbf{in} \textbf{$M$}
\State $P_{v,t,h} \gets \sum_m^M C_m$
\State $P_{T,t,h} \gets P_{b,t-1,h} + P_{v,t,h-L}$ \\
\Return $P_{v,t,h}$, $P_{T,t,h}$
\end{algorithmic}
\end{algorithm}

Specifically, the algorithm takes inputs such as an image, a vector of target vehicles types, the lag between camera-converted and sensor PM2.5 index and  base PM2.5 concentration of other emission sources of the previous day, then it outputs the PM2.5 concentration of emission from vehicles and the aligned total PM2.5 concentration in the current day $t$ of an hour $h$. The latter two arguments are re-calculated once at the end of a day based on Algorithm \ref{alg:basePMlag} in other to prepare for the next calculation round of Algorithm \ref{alg:vehicle2PM}. In Algorithm \ref{alg:vehicle2PM}, there are some intermediate variables, which have not been specified, such as  $\{\textbf{O}\}$ is the vector of extracted foreground objects, $\{\textbf{M}_i: \textbf{O}_j\}$ stands for a dictionary which list desired vehicle types and their list of belonging objects, $N_m, E_m, C_m$ are the previously introduced factors which stand for number of vehicles of type $m$, the corresponding preliminary PM2.5 emission and the target PM2.5 concentration over a given area and time, respectively. 

\begin{algorithm}
\caption{PM2.5 concentration base from other sources than vehicles and lag between camera-converted and sensor PM2.5 index calculation algorithm}\label{alg:basePMlag}
\begin{algorithmic}
\Require $P_{v,0\dots23}$: PM2.5 concentration of emission from vehicles spreading over 24 hours, \\ $\quad \quad \quad \ P_{s,0\dots23}$: PM2.5 concentration of emission from an air sensor spreading over 24 hours
\Ensure  $lag$: lag between camera-converted and sensor PM2.5 index \\ $\quad \quad \quad P_{b,0\dots23}$: base PM2.5 concentration of other emission sources spreading over 24 hours

\State $\{lag\} \gets \textbf{Cross-correlation}(P_{v,0\dots23},P_{s,0\dots23})$
\State $lag_{max} \gets \textbf{max}(\{lag_i\})$
\State $P_{b,0\dots23} \gets P_{s,0\dots23} - P_{v,0-lag_{max}\dots23-lag_{max}}$ \\
\Return $lag$, $P_{b,0\dots23}$
\end{algorithmic}
\end{algorithm}

\section{Sheaf-theoretic self-filtering network modeling}
\label{sec:sheafmodel}
Several tracking methods have been proposed to solve the problem of integrating heterogeneous information. The most common method is aggregation or pooling, which involves evaluating all available data points and combining them into a single value. However, this method can be inaccurate when the data points are highly dissimilar. Instead of pooling data, tracking experts have developed models that track data points individually and then fuse them together at the end. For example, in an election where candidates run on different platforms, it's difficult to integrate the different inputs into an accurate model. In contrast, a sheaf-based tracking model makes sense when multiple sources are tracking different aspects of a particular situation. A sheaf-based tracking model is an aggregation model that tracks a specific quantity instead of individual data points. One method for doing so is to create a sheaf from all the available data points and integrate all the sheaves together to calculate the total amount of heterogeneous information. The sheaf-based integration method is much more accurate than other aggregation models since it allows for individual evaluation of each piece of data within a sheaf. To evaluate a point in a sheaf, one needs only to locate the corresponding point in space; this is much easier than locating a specific data point within a pooling model.

A sheaf is a data structure for storing information over a topological space. The topological space specifies the relationships between the sensors, but the sheaf operates on the unprocessed input data, mapping all the sources into a common comparison framework. Abstract simplicial complex (ASC) is a discrete mathematical entity reflecting not only the available sensors as vertices (in this case, the four physical sensors including two cameras and two air quality sensors) but also their multi-way interactions as higher-order faces. Where the ASC forms the base of the sheaf, the faces of the stalks hold the recorded information. The sheaf model is completed by specifying restriction functions that model interactions between sensor combinations and data. In this context, we can define assignments as recorded readings, sections as assignments that are all consistent based on the sheaf model, and partial assignments and sections over a subset of sensors. 

Then, we deliver the following strategies for sheaf modeling. In particular, when global or partial sections imply totally consistent data, we add \textit{consistency structures} to describe data that are only partially consistent. \textit{Consistency structures} instantiated for the sensor nodes model specifically take the shape of n-way standard deviations, although being specified totally broadly. In turn, \textit{consistency structures} allow us to create approximation sections that can quantify the degree of sensor consistency. The \textit{consistency radius} is a native global measure of the uncertainty among the sensors present in any given reading. In addition, \textit{consistency filtration} \cite{robinson2017sheaves} gives a comprehensive description of the contributions of individual sensors and sensor combinations to the total uncertainty. 

\subsection{Sheaf simplicial constructions}
We begin by defining an abstract simplicial complex, the sort of topological space utilized to describe our sensor network. Typically, specifying all of the simplices in a simplicial complex is tedious. Instead, it is considerably more practical to provide a generating set X of subsets of the vertex set. Abstract simplicial complex formed by X is the unique smallest simplicial complex comprising the generating set, and formally defined, according to \cite{joslyn2020sheaf}, as follows.

\begin{definition}[Abstract Simplicial Complex]
An abstract simplicial complex $X$ on a set $V_X$ is a set of subsets of $V_X$, where $\sigma \in X$ and $\gamma \subseteq \sigma$, then $\gamma \in X$. Each $\sigma \in X$ is a simplex of $X$, whereas each $V_X$ element is a vertex of $X$. Every subset $\gamma$ of a simplex $\sigma $ is referred to as a \textbf{face} of $\sigma $. More generally, each $\sigma \in X$ with d+1 components is referred to as a \textbf{d-face} of $X$, with \textbf{d} standing for its dimension. \textbf{Vertices} are zero-dimensional faces (singleton subsets of $V_X$), while \textbf{edges} are one-dimensional faces. 
\end{definition}

The following describes how an abstract simplicial complex may be used to represent the links inside a sensor network. Consider the base set $V_X$ to be the network's collection of sensors, and consider $X$ to include every collection of sensors that measure the same amount. According to the sensor network description in Section \ref{section:network}, the following symbols have been assigned to the four sensors, comprising two camera nodes and two dust particle sensory nodes, that are coupled to track the local air quality signal: $C_1$, $C_2$, $S_1$ and $S_2$, respectively. 

As all four sensors point to one final set of information, the PM2.5 index, we define $U = C 1, C 2, S 1, S 2$ as our base sensor set, where $C = C 1, C 2 $ consists of a set of cameras that count and classify the traffics and $S = S 1, S 2 $ comprises two air sensor devices that measure the PM2.5 index of the air. In this way, our air-monitoring network's ASC $X$ has a total of ten faces, counting both $C$ and $S$ and their subsets: 

\begin{equation}
\begin{split}
X = \{ \{C_1,C_2\}, \{S_1,S_2\}, \\
    \{C_1,S_1\}, \{C_2,S_2\},\{C_1,S_2\}, \{C_2,S_1\}, \\
    \{C_1\}, \{S_1\}, \{C_2\}, \{S_2\} \}
\end{split}
\end{equation}

\begin{figure}[t!]
  \centering
  \includegraphics[width=0.75\linewidth]{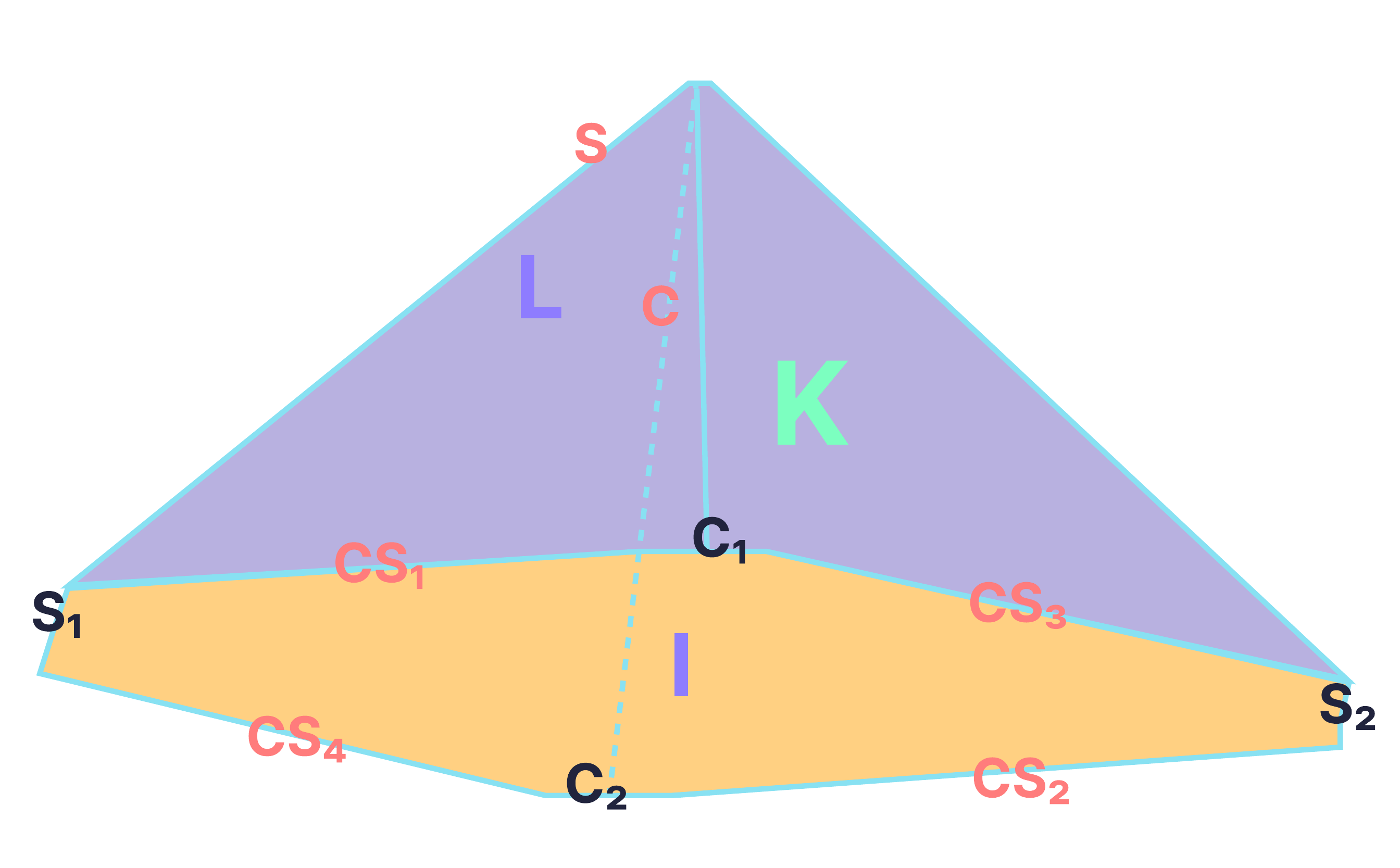}
  \caption{Simplicial complex of our network of air monitoring sensors.}
  \label{fig:sheafbase}
\end{figure}

The remaining pairwise sensor interaction faces are denoted as $CS_1 = \{C_1,S_1\}$, $CS_2 = \{C_2,S_2\}$, $CS_3 = \{C_1,S_2\}$, $CS_4 = \{C_2,S_1\}$; hence, the whole sheaf diagram is depicted in Figure \ref{fig:sheafbase}, where the highest dimensional face (the final air PM2.5 index \textit{K}) and all sub-faces are indicated. The sensors are depicted as black singleton faces, whereas the higher dimensional faces are depicted in red. In addition, the solid rectangle symbolizes the three-way interaction \textit{I}. 

\begin{figure}[h!]
  \centering
  \includegraphics[width=0.75\linewidth]{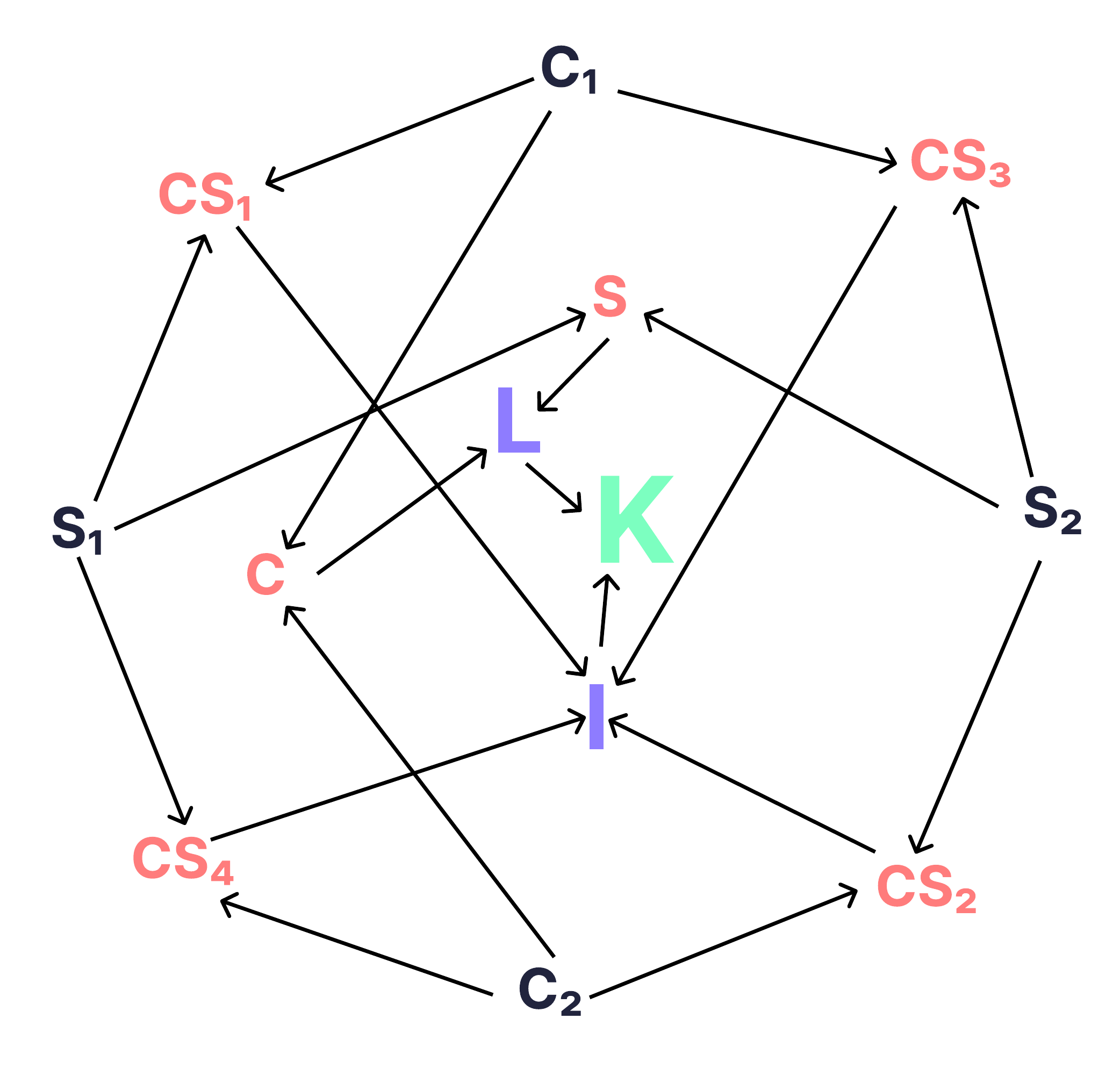}
  \caption{Corresponding attachment diagram of our network of air monitoring sensors.}
  \label{fig:sheafbasedag}
\end{figure}

The ASC implies a topological space that expresses multidirectional interactions. Each d-dimensional face is a hyper-tetrahedron held together by sensor interactions. This fundamental structure has one connected component and no "open loops." Depending on the number of observables and their configuration, this structure can become arbitrarily complicated, with high-order faces and complex connections that may include open loops or voids. Sheaf theory represents all interactions automatically. Figure \ref{fig:sheafbasedag} depicts the attachment diagram associated with the ASC. This is a directed acyclic graph (DAG) in which nodes are the faces of the ASC and directed edges lead upward from a face to its associated face (co-face) of higher dimension.

Accordingly, a \textit{sheaf} is a compatible assignment of data to each face and defined, according to \cite{joslyn2020sheaf}, as follows
\begin{definition}[A Sheaf on an Abstract Simplicial Complex]
A \textbf{sheaf} of sets, denoted by the notation $\mathcal{S}$, on an abstract simplicial complex $\mathcal{X}$ is made up of the assignment of (1) a set $\mathcal{S}(\delta)$ to each face $\delta$ of $\mathcal{X}$ (called the \textbf{stalk} at $\delta$), and (2) the restriction map from $\gamma$ to $\delta$ denoted as a function $\mathcal{S}(\gamma \leadsto \delta) : \mathcal{S}(\gamma) \leadsto \mathcal{S}(\delta)$ such that
$$
\mathcal{S}(\gamma \leadsto \theta) \circ \mathcal{S}(\gamma \leadsto \delta) = \mathcal{S}(\gamma \leadsto \theta), \quad iff \quad \gamma \leadsto \delta \leadsto \theta
$$
\end{definition}

In other words, a \textit{sheaf of vector spaces} allocates a vector space to each face and a linear map to each attachment in a similar manner. \textit{A sheaf of groups} further assigns a group to each face and a group homomorphism to each attachment. Concretely, the stalk above a face is the location where the data connected with that face resides. Restriction functions create the grounds on which data that interacts with one another can be considered to be consistent or inconsistent. 

\begin{figure}[h!]
  \centering
  \includegraphics[width=0.75\linewidth]{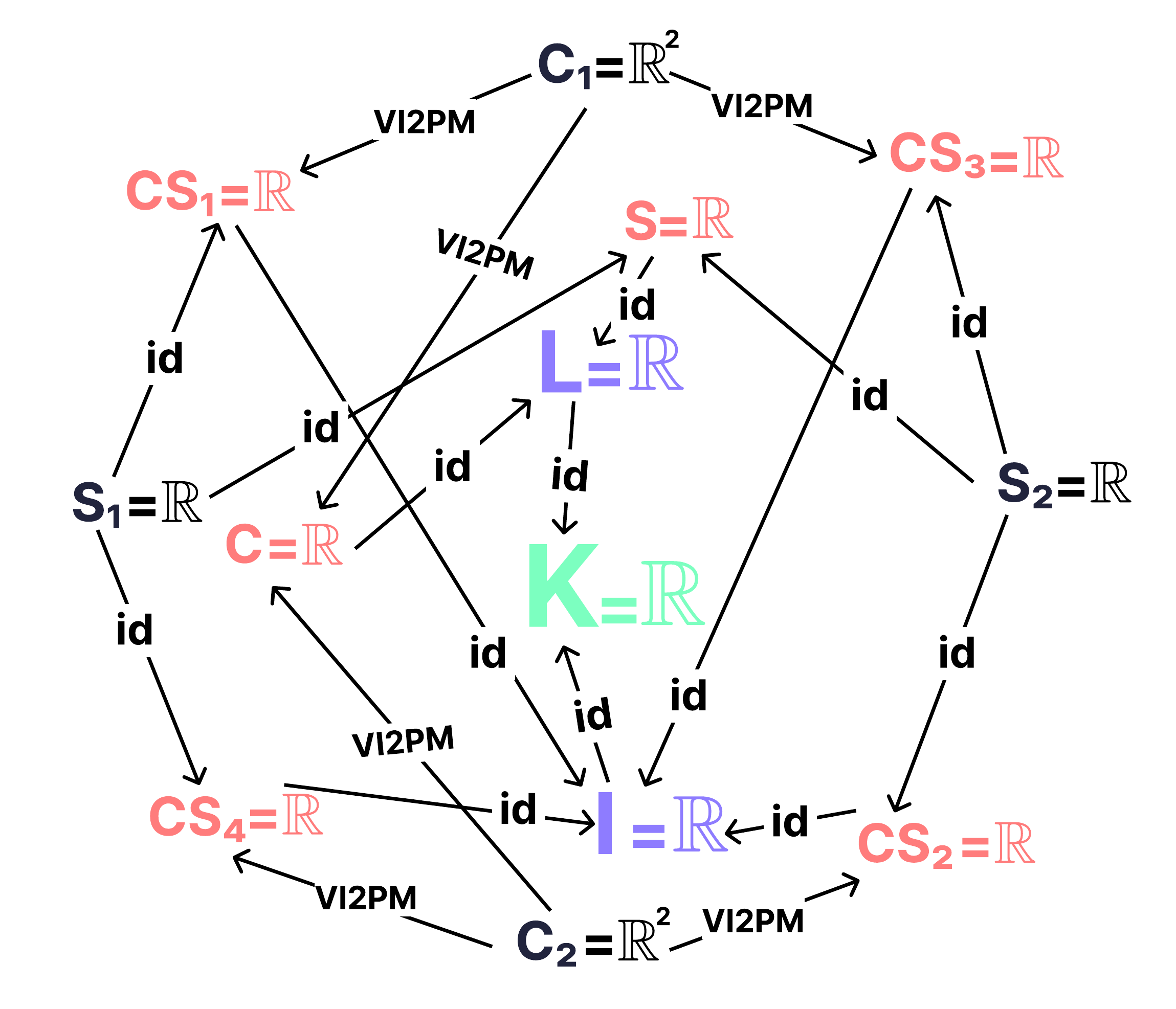}
  \caption{Sheaf model of the air monitoring network. GB stands for guide book, which is used to translate from the vehicle numbers to PM2.5 - Refer to Section \ref{section:network} for more details. The other abbreviation, id, stands for "identity", which then refers to the identity transformation, or in the sheaf language, the identity restriction map.}
  \label{fig:sheafdag}
\end{figure}

For our air monitoring network in particular, we have two cameras, each of which encodes vehicle counts over time from the input video stream; hence, the stalks over the vertices $C 1$ and $C 2$ are $\mathbb{R}$. Similarly, the stalks above the vertices of the sensory nodes are also $\mathbb{R}$ since they monitor the PM2.5 levels directly. Now that the stalks for the vertices have been established, the stalks for the higher order faces and the restriction maps must be determined. However, the data formats are not compatible with one another. Therefore, in order to compare these measures, the information should initially be converted into standard units along the edges, and then it should be transmitted to the rectangular face. We decided to use PM2.5 as our common coordinate system because most of the sensors report their readings in these formats. Then $\mathbb{R}$ will serve as the pillar for the higher order sides of the rectangle. We make use of a guide book, which is discussed in more detail in the aforementioned Section \ref{section:network}, in order to nonlinearly translate the vehicle counts from the camera vertices to the common unit in all of the other faces. Then, the attachment diagram in Fig. \ref{fig:sheafbasedag} becomes our sheaf model for the air monitoring network in Fig. \ref{fig:sheafdag}

The sheaf model describes the temporal agreement of sensors as described in the following. At a given time $t$, each vertex is allocated the most recent sensor reading, a data point from its stalk space. These measurements are subsequently sent to the higher order faces for comparison through the restriction maps. If the two measurements obtained by an edge are identical, this single value is assigned to that edge, and the algorithm proceeds. Towards this process, an agreement happens when a single value is set for faces of the same kind, e.g. Cameras (the same value is set for $C$, $C_1$ and $C_2$). When there is the potential for a complete assignment, we refer to that as a \textit{global section}. The possibility of not agreeing is always present, which is where the idea of an \textit{assignment} comes from. This paves a way for the following definition, according to \cite{joslyn2020sheaf}.

\begin{definition}[Assignment and global section]
Let $\mathcal{S}$ be a sheaf on an abstract simplicial complex \textit{X}. A function $\gamma: X \rightarrow \Pi_{x \in X}\mathcal{S}(x)$ which yields a value $\gamma(x) \in \mathcal{S}(x)$ to each face $x \in X$ is defined as an \textbf{assignment}. Then, the definition of a \textbf{partial assignment} is bounded by that of an  \textbf{assignment}: a function $\alpha: X' \rightarrow \Pi_{x \in X'}\mathcal{S}(x)$ which yields a value $\gamma(x) \in \mathcal{S}(x)$ to each face $x \in X' \subset X$. An assignment $s$ is referred to as a \textbf{global section} if for each inclusion $x \leadsto \beta$ of faces, $\mathcal{S}(x \leadsto \beta)(s(x)) = s(\beta)$.
\end{definition}

At a particular moment, a global section of our tracking model corresponds to the sensor data agreeing simultaneously. The equality condition of a global section may be too restrictive for certain applications, such as our tracking model. 

\subsection{Loosening Section constraints with Consistency Structures}
The constraint that sensor data on the vertices match, on going through limitations, is loosened up by consistency structures, which instead require that the sensor data simply agree with one another. Consistency structures loosen global section constraints because they allow for the formation of global sections from compatible local sections, rather than requiring all local sections to agree on a single global section. In which, a boolean function is applied to each individual face in order to determine the level of agreement. Together, a sheaf and a collection of these boolean functions constitute what is known as a consistency structure. Formally, a consistency structure is defined, according to \cite{joslyn2020sheaf}, as follows.

\begin{definition}[Consistency structures]
A consistency structure is defined as a triple (X, $\mathcal{S}$, \textbf{C}) in which X is an abstract simplicial complex, $\mathcal{S}$ is a sheaf over X and \textbf{C} is the assignment to each non-vertex d-face $\beta \in X, d>0$, of the following function:
$$
\textbf{C}_{\beta}: \left(\left( \mathcal{S}(\beta) \atop dim \beta + 1 \right)\right) \to \{0,1\}
$$
, where the double bracket pairs derive the set of subsets with length $\beta + 1$ over $\mathcal{S}(\beta)$.
\end{definition}

The multiset in the domain of $C_{\beta}$ represents the numerous sheaf values to be compared for all vertices infringing on a non-vertex face $\beta$, while the codomain $\{0, 1\}$ denotes whether or not they match "sufficiently." Formally, a \textbf{conventional consistency structures} for a sheaf $\mathcal{S}$ applies an equality test to each face that is not a vertex $\beta = \{v_1, v_2, \dots, v_k\}$, according to \cite{joslyn2020sheaf}
\begin{equation}
\textbf{C}_{\beta}([z_1, z_2, \dots, z_k]) = 
\begin{cases}
1, \quad if \; z_1 = z_2 = \dots = z_k \\
0, \quad otherwise
\end{cases}
\end{equation}
, where square bracket pair stands for the multiset and $z_j = \mathcal{S}(\{v_i\} \leadsto \beta)(s(v_j))$. A consistency structure expands the equality criterion inherently accessible in a sheaf to include classes of values that are regarded as equivalent. Beyond that, each stalk in our tracking model is a metric space. Thus, we may use the natural metric to determine if two points are only "near enough" as opposed to entirely congruent or comparable. Using $\epsilon$ to represent the amount of mistake present or acceptable in an assignment, we construct the \textbf{$\epsilon$-approximate consistency structure} for a sheaf $\mathcal{S}$ a nd each of the non-vertex $\beta = \{v_1, v_2, \dots, v_k\}, k>1$, according to \cite{joslyn2020sheaf}, as follows.
\begin{equation}
\textbf{C}_{\beta}([z_1, z_2, \dots, z_k]) = 
\begin{cases}
1, \quad if \; \delta([z_1, z_2, \dots, z_k]) \leq \epsilon \\
0, \quad otherwise
\end{cases}
\end{equation}
, where the function $\delta$ measures the consistency as a general spread of the multivariate data. Specifically, it is defined as, according to \cite{joslyn2020sheaf}:
\begin{equation}
\label{eq:consistencymeasure}
\delta(Y) = \sqrt{\frac{1}{|Y|} \textbf{Tr}(\Sigma_Y)}
\end{equation}
, in which $\Sigma_Y$ is the covariance matrix of the multidimensional data $Y$.

Then, a consistency structure's equivalent of a global section for sheaves is known as a pseudosection. A pseudosection assignment $s$ guarantees that, for each non-vertex face $\beta$, (i) the restrictions of its vertices to the face are sufficient, and (ii) the value given to the face is consistent with the restricted vertices. Formally, the pseudosection definition can be described, according to \cite{joslyn2020sheaf}, as follows:
\begin{definition}[Pseudosection]
For each non-vertex $\beta = \{v_1,v_2,\dots,v_k\}$, an assignment $s \in \Pi_{x \in X} \mathcal{S}(x)$ is defined as a $(X,\mathcal{S},\textbf{C})$-pseudosection if 

$
i. \quad \textbf{C}_{\beta}([\mathcal{S}(\{v_i\} \leadsto \beta)s(\{v_i\}): i = 1,\dots,k]) = 1, \ and
$

$
ii. \quad \textbf{C}_{\beta}([\mathcal{S}(\{v_i\} \leadsto \beta)s(\{v_i\}): i = 1,\dots,j-1,j+1,\dots,k]) \cup s(\{v_i\})) = 1,\ for\ j\ in\ 1,2,\dots,k.
$
\end{definition}

Hence, given this, the pseudosections of a conventional consistency structure are global sections of its associated sheaf because when \textbf{C} is the conventional consistency structure, the condition for an assignment for each non-vertex face $\beta = \{v_1, v_2, \dots,v_k\}$ to become a pseudosection becomes
$s(\beta) = \mathcal{S}(v_1 \leadsto \beta) s(\{v_1\}) = \mathcal{S}(v_2 \leadsto \beta) s(\{v_2\}) = \dots = \mathcal{S}(v_k \leadsto \beta) s(\{v_k\})$. Whereas, if \textbf{C} is an $\epsilon$-approximate consistency structure, pseudosections for each non-vertex face $\beta = \{v_1, v_2, \dots,v_k\}$ are defined as
\begin{equation}
\begin{cases}
\quad \delta([\mathcal{S}(\{v_i\} \leadsto \beta)s(\{v_i\}): i = 1,\dots,k]) \leq \epsilon \\
\quad \delta([\mathcal{S}(\{v_i\} \leadsto \beta)s(\{v_i\}): i = 1,\dots,j-1,j+1,\dots,k]) \cup s(\{v_i\})) \leq \epsilon, \\for\ j\ in\ 1,2,\dots,k.
\end{cases}
\end{equation}

A pseudosection of our tracking model's $\epsilon$-approximate consistency structure assigns the PM2.5 signals such that the "spread" of all measurements ascribed to each face is constrained by $\epsilon$. According to the following theorem as in \cite{joslyn2020sheaf}, the lowest $\epsilon$ for which a pseudosection occurs is solely defined by restricting the images of the vertices. 

\begin{theorem}[Consistency Radius]
Let $\mathcal{S}$ be a sheaf on an abstract simplicial complex $X$ where each stalk is a metric space and $s$ be an assignment such that it belongs to the set $\Pi_{x\in X}\mathcal{S}(x)$. The existed minimum $\epsilon$ such that $s$ is a pseudosection of the $\epsilon$-approximate consistency structure $(X,\mathcal{S},\textbf{C})$
$$
\epsilon^* = \max\limits_{\beta \in X \setminus \{\{v\}: v \in V \}}\delta([\mathcal{S}(w \leadsto \beta) s(\{w\}): w \in \beta])
$$
, where $V$ is the set of vertices of the sheaf and the proof is stated in \cite{joslyn2020sheaf}.
\end{theorem} 

\begin{lemma}[Consistency Radius]
Should a set of real numbers $Z = \{z_1, z_2, \dots, z_k\}$ that has a mean of $\mu_Z$, $\forall z \in Z, \delta(Y_z) \leq \delta(Z)$, where $Y_z = Z \setminus z \cup \mu_Z$.
\end{lemma}

The consistency radius definition naturally leads to the concern of how much the consistency of the network system is among the nodes when $\epsilon$ ranging from the lowest value to the defined consistency radius. Data is received from several signal sources, some of which may or may not agree. Thus, it is desirable to select maximally consistent network sections, which pave the way for other essential metrics such as consistency filtration and cover measures. According to \cite{joslyn2020sheaf}, the subsequent theorem and its corresponding lemma demonstrate how to associate a unique collection of maximally consistent subcomplexes with any vertex assignment.

\begin{theorem}[Maximally Consistent Subcomplexes]
\label{theorem2}
Let a consistency structure $(X,\mathcal{S},\textbf{C})$ be with a sheaf partial assignment $s$ on vertices $\mathcal{U}$. There exists a unique group of subsets $\{W_i\}$ of $\mathcal{U}$ that leads to subcomplexes $\{X_{W_i}\}$ of X with the following characteristics: \\
i. The assignment $s$ is consistent within each $X_{W_i}$, and any subcomplex where $s$ is consistent also has at least one $X_{W_i}$ as a supercomplex. \\
ii. $\bigcup star(X_{W_i})$ is a cover of $X$, where $star(X_{W_i})$ is defined as the set of faces containing $X_{W_i}$, formally $star(X_{W_i}) = \{k \in X: X_{W_i} \subseteq k \}$.
\end{theorem}

\begin{lemma}[Maximally Consistent Subcomplexes]
Let a consistency structure $(X,\mathcal{S},\textbf{C})$ be with a sheaf partial assignment $s$ on $\mathcal{U}$. if for some non-vertex face $\beta in X$ such that $C_{\beta}(\mathcal{S}(\{v\} \leadsto \beta)s(\{v\})) = 0$, there exists subsets $\{W_i\}$ of $\mathcal{U}$ which leads to the subcomplexes $\{X_{W_i}\}$ of $X$ such that: \\
i. $\beta \notin \{X_{W_i}\}$ for all $i$ \\
ii. The assignment $s$ is consistent within each $X_{W_i}$, and any subcomplex where $s$ is consistent also has at least one $X_{W_i}$ as a supercomplex. \\
iii. $\bigcup star(X_{W_i})$ is a cover of $X$, where $star(X_{W_i})$ is defined as the set of faces containing $X_{W_i}$, formally $star(X_{W_i}) = \{k \in X: X_{W_i} \subseteq k \}$.
\end{lemma}

Theorem \ref{theorem2} provides a maximal vertex cover for our tracking model when given a collection of sensor readings and a value for $\epsilon$ in order to ensure that each associated subcomplex is approximately consistent within a specified error margin. In order to define a measure for the vertex cover associated with a set of maximal consistent subcomplexes, we treat the set of covers as a graded poset, and use the rank function of the poset as a measure.

\begin{definition}[Graded posets and rank function]
Let $\mathcal{P} = \langle P, \trianglelefteq \rangle $ be a poset, it is \textbf{graded} if there exists a \textbf{rank} function $r: P \rightarrow \mathbb{N} \cup \{0\}$ such that $r(s) = 0$ if $s$ is a minimal element of $P$ as well as $r(j) = r(k) + 1$ if $j \lhd k$ in $P$. $s$ is said to have rank $i$, if $r(s) = i$ and the maximum rank, which is defined as $\max\limits_{p\in P}\{r(p)\}$, stands for the rank of $\mathcal{P}$.
\end{definition}

Finally, the rank function can be used to quantify the vertex cover associated with a set of maximally consistent subcomplexes, thanks to the detailed proof in \cite{joslyn2020sheaf}. The following definition specifies how this is proceeded.

\begin{definition}[Graded posets and rank function]
Let a consistency structure $(X,\mathcal{S},\textbf{C})$ that has the number of vertices $|V| = n$ and $\mathcal{A} = \{W_i\}$ be a vertex cover obtained from Theorem \ref{theorem2}. The measure of such vertex cover is defined as 
$$
\bar{r} (\mathcal{A}) = |\downarrow \mathcal{A}| - (n+1)
$$

Note that $0 \leq \bar{r}(\mathcal{A}) \leq 2^n - (n + 1)$, and a bigger number suggests subcomplexes with more consistency and given $\downarrow\mathcal{A}_1 \subseteq \downarrow\mathcal{A}_2$ entails $\bar{r}(\mathcal{A}_1) \leq \bar{r}(\mathcal{A}_2)$.
\end{definition}

\subsection{Consistency Filtrations}
Thus far, we have defined a simplicial complex $X$, a sheaf $\mathcal{S}$ on the simplicial complex, a partial assignment $s$ to the vertices of $X$ and $\epsilon$-approximate consistency structure $(X, \mathcal{S}, \textbf{C}_{\epsilon})$. Such elements help build up a consistency filtration by varying $\epsilon$ and thanks to the Theorem \ref{theorem2} that the set of maximal consistent subcomplexes could be obtained, the filtration of vertex covers corresponding to landmarks ranging over $\epsilon$ values listed as $\epsilon_0 = 0 < \epsilon_1 < \dots < \epsilon_{t-1} < \epsilon_{t} = \epsilon^*$ - the consistency radius, which is the smallest possible $\epsilon$ such that it makes the assignment $s$ a $(X, \mathcal{S}, \textbf{C}_{\epsilon})$-pseudosection. The sequence of vertex covers, $C_0 \trianglelefteq C_1 \trianglelefteq \dots \trianglelefteq C_{t-1} \trianglelefteq C_{t}$, can be refined such that each set consists of subcomplexes whose union forms $X$. We can also compute the sequence of cover measures, $p_0 \trianglelefteq p_1 \trianglelefteq \dots \trianglelefteq p_{t-1} \trianglelefteq p_{t}$, for each set of covers, $C_i$. The consistency filtration is a method for evaluating the consistency of a group of sensors. If the distance between two adjacent landmark values, $\epsilon_i$ and $\epsilon_{i+1}$, is significantly greater than the other distances, this indicates that there is a disagreement between at least two groups of sensors. By comparing the covers $C_i$ and $C_{i+1}$, it is easy to see which sensors are causing the disagreement.

\section{Integrated algorithm}
The algorithm we will define in this section is an important step in understanding the behavior of our system over time. By using the information from the nodes at a given time, we can determine the consistency filtration of the system, which is a measure of the coherence and stability of the system at that time. This allows us to track the evolution of the system and identify any potential issues or instability.

The system populates the data to each node in the set $\{C_1, C_2, S_1, S_2\}$ asynchronously, once every 10 minutes for camera nodes and once every 15 seconds for the air sensor nodes. The data is first retrieved by the lowest-ranking face of the sheaf and then distributed, or "lifted up," to higher faces level by level until it reaches the highest-ranking face. Specifically to the sheaf model, described in Fig. \ref{fig:sheafdag}, the second level is the set of edges $\{CS_1, CS_2, CS_3, CS_4, C, S\}$, and the highest level is $K$. Please note that $K$ is illustrated as a 3-dimensional space formed by 2-dimensional surfaces $I$ and $L$. Each time the data is lifted in the sheaf model, the corresponding conventional consistency structure of the sheaf model must have pairs of adjacent nodes/edges that are strictly equal. This means that the values of all high-level faces in the sheaf model must be zero, forming a pseudosection of the conventional consistency structure, or a global section of the sheaf model. However, as mentioned above, achieving a global section in real-world applications in unrealistic because we would never have perfect working system; thus, we would use $\epsilon$-approximate consistency structure instead to allow a fraction of errors among the faces. Then, the propagated values along the faces in the consistency structure is described in Eq. \ref{eq:consistencymeasure}.

Thanks to the definition of consistency radius and consistency filtration, we could know which group of sensors is consistent to each other. Similar to the work of Joslyn et al. \cite{joslyn2020sheaf}, we could use the consistency measure to detect the sensing nodes at fault and decide to isolate their results before checking up their status. In summary, the whole aforementioned process could be determined through the following algorithm, designed specifically for the sheaf model in Fig. \ref{fig:sheafdag}.

\begin{algorithm}
\caption{Self-consistency management from heterogeneous input sources algorithm}\label{alg:selfConsistency}
\begin{algorithmic}
\Require $\mathcal{S}$: sheaf model, $\{C_1, C_2, S_1, S_2\}$: vertices set
\Ensure $\epsilon_c$: consistency threshold, \\ $\quad \quad \quad V_c$: value of the face at $\epsilon_c$, \\ $\quad \quad \quad S_c$: a set of consistent sensors
\State \textbf{$0\mhyphen faces$} $\gets \{C_1, C_2, S_1, S_2\}$ \Comment{Original vertices, or $0\mhyphen faces$}
\State $CS_1() \gets \{\mathcal{S}(C_1 \leadsto 1\mhyphen faces) s(\{C_1\}), \mathcal{S}(S_1 \leadsto 1\mhyphen faces) s(\{S_1\})\}$ 
\State $CS_2() \gets \{\mathcal{S}(C_2 \leadsto 1\mhyphen faces) s(\{C_2\}), \mathcal{S}(S_2 \leadsto 1\mhyphen faces) s(\{S_2\})\}$
\State $CS_3() \gets \{\mathcal{S}(C_1 \leadsto 1\mhyphen faces) s(\{C_1\}), \mathcal{S}(S_2 \leadsto 1\mhyphen faces) s(\{S_2\})\}$
\State $CS_4() \gets \{\mathcal{S}(C_2 \leadsto 1\mhyphen faces) s(\{C_2\}), \mathcal{S}(S_1 \leadsto 1\mhyphen faces) s(\{S_1\})\}$
\State $C() \gets \{\mathcal{S}(C_1 \leadsto 1\mhyphen faces) s(\{C_1\}), \mathcal{S}(C_2 \leadsto 1\mhyphen faces) s(\{C_2\})\}$
\State $S() \gets \{\mathcal{S}(S_1 \leadsto 1\mhyphen faces) s(\{S_1\}), \mathcal{S}(S_2 \leadsto 1\mhyphen faces) s(\{S_2\})\}$
\State \textbf{$1\mhyphen faces$} $\gets \{CS_1, CS_2, CS_3, CS_4, C, S \}$ \Comment{Configuring $1\mhyphen faces$}
\State $I \gets \{\mathcal{S}(CS_1 \leadsto 2\mhyphen faces) s(\{CS_1\}), \mathcal{S}(CS_2 \leadsto 2\mhyphen faces) s(\{CS_2\}),$
$\mathcal{S}(CS_3 \leadsto 2\mhyphen faces) s(\{CS_3\}), \mathcal{S}(CS_4 \leadsto 2\mhyphen faces) s(\{CS_4\})\}$
\State $L \gets \{\mathcal{S}(C \leadsto 2\mhyphen faces) s(\{C\}), \mathcal{S}(S \leadsto 2\mhyphen faces) s(\{S\})\}$ 
\State \textbf{$2\mhyphen faces$} $\gets \{I,L\}$ \Comment{Configuring \textbf{$2\mhyphen faces$}}
\State $K \gets \{\mathcal{S}(I \leadsto 2\mhyphen faces) s(\{I\}), \mathcal{S}(L \leadsto 2\mhyphen faces) s(\{L\})\}$
\State \textbf{$3\mhyphen faces$} $ \gets \{K\}$ \Comment{Configuring \textbf{$3\mhyphen faces$}}
\State $V \gets \{\}$  \Comment{Initializing an empty dictionary that stores complexes' values}
\State $\epsilon \gets \{\}$ \Comment{Initializing an empty dictionary that stores complexes' corresponding $\epsilon$ threshold}
\While{$d \leq 3$}
\State $\Omega \gets d\mhyphen faces$
\If{$d > 0$}
\For{$\omega \in \Omega$}
\State $\{\omega_1, \omega_2\, \dots, \omega_n \} \gets \omega()$
\State $V[\omega] \gets \{\omega_1, \omega_2, \dots, \omega_n \}$
\State $\epsilon[\omega] \gets \delta(\{\omega_1, \omega_2, \dots, \omega_n \})$ \Comment{$\delta(.)$ is determined in Eq. \ref{eq:consistencymeasure}}
\EndFor
\Else 
\State $V[\omega] \gets s(\omega)$ \Comment{Initializing with partial assignment}
\State $\epsilon[\omega] \gets 0$
\EndIf
\EndWhile
\State $S_c \gets []$
\State $\epsilon_c \gets 0$
\State $V_c \gets 0$
\For{$\omega \in \Omega$}
\If{$\epsilon[\omega] \leq mean(\epsilon) + 0.5*std(\epsilon)$} \Comment{Checking if the consistency threshold belong to such part of the filtration}
\State $S_c.append(\omega)$
\State $\epsilon_c \gets \epsilon[\omega]$
\State $V_c \gets V[\omega]$
\EndIf
\EndFor \\
\Return $\epsilon_c$, $V_c$, $S_c$
\end{algorithmic}
\end{algorithm}

The algorithm appears to be using data aggregation to evaluate the consistency of our sheaf model through the mean operator. However, this is not a simple aggregation approach. This can be shown through two points: (i) our sheaf model uses data from various heterogeneous sensors taken at different times, and (ii) the mean operator is not applied to all nodes, but is instead used hierarchically across different levels of complexes and is only applied to the most consistent sensors, which is defined by the being less than or equalled to the mean + half a standard deviation of the filtration distribution. These points are illustrated in more details in our toy examples in Section \ref{sec:toyEx}.

\section{Toy Examples}
\label{sec:toyEx}

In this section, we will introduce a set of toy examples to demonstrate how sheaf modeling works using simulated sensor measurements and generated data. To provide a concrete illustration of the concepts involved, we will use simple examples that can be easily visualized and understood. The goal of this example is to provide a gentle introduction to sheaf modeling and to give a sense of how it can be applied in practice. By working through this section, we hope to give the reader a better understanding of the main ideas behind sheaf modeling and how it can be used to analyze and interpret data.

\subsection{Example 1: Sheaf Global Section or Consistency Structure Pseudo-section}
In this example, we will showcase the behavior of our built sheaf when it has a global section, or pseudo-section if we take into account the consistency structure of the sheaf. This means that all of the assignments to the vertices and faces of the sheaf perfectly match together, resulting in a cohesive and well-defined global structure. By examining this scenario, we can gain insight into how the sheaf behaves when all of the local assignments fit together seamlessly and how it can be used to represent and analyze data in this context. 

Specifically, we code up our sensor network sheaf model, illustrated in Fig. \ref{fig:sheafdag}. We use the Python \textit{networkx} package to visualize the sheaf we have built by the function we list in the Supplementary Section \ref{code:sheaf} in Fig. \ref{fig:sheafpydag}.

\begin{figure}[h!]
  \centering
  \includegraphics[width=0.75\linewidth]{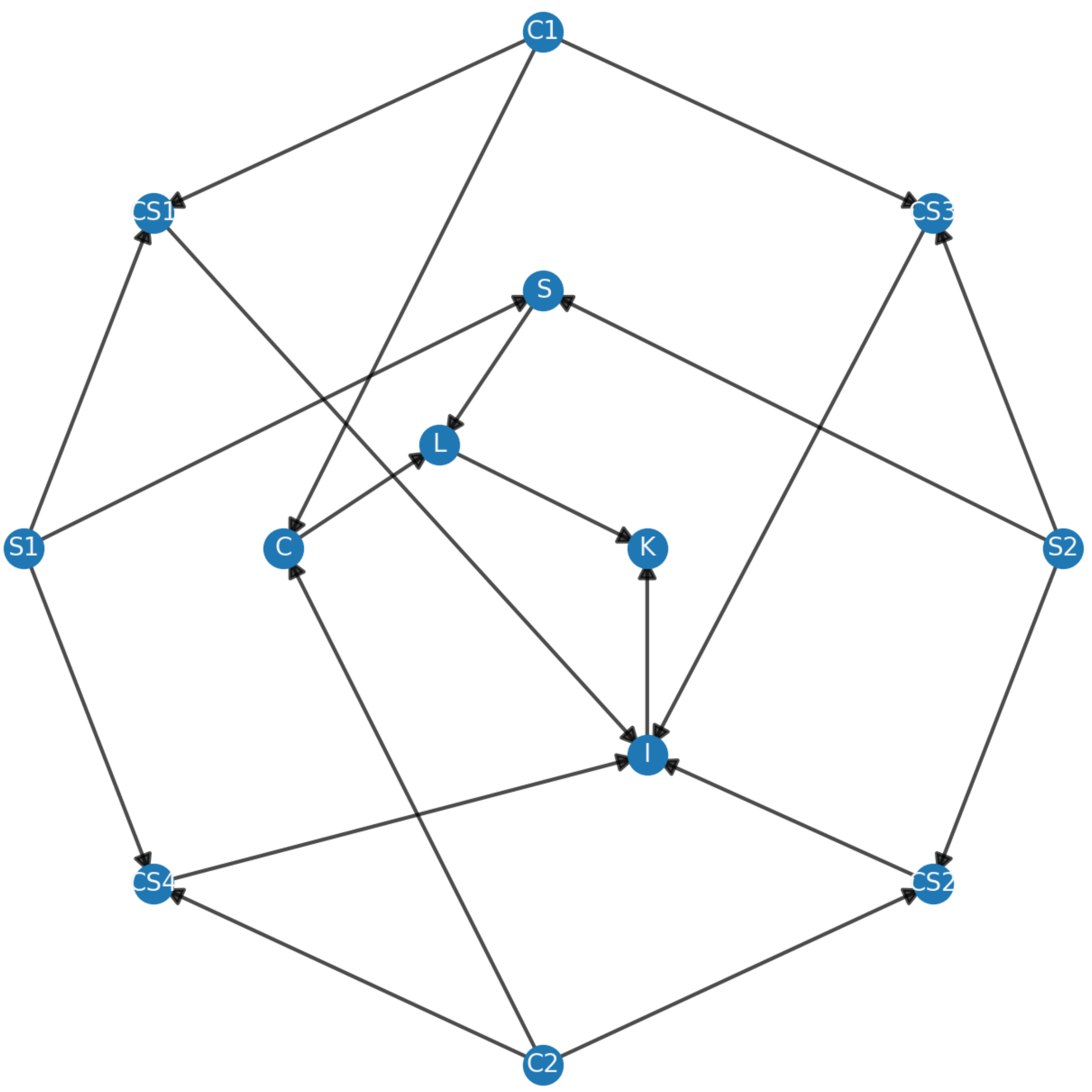}
  \caption{Sheaf model of the air monitoring network, drawn by the Python \textit{networkx} package.}
  \label{fig:sheafpydag}
\end{figure}

\begin{table}\centering
\caption{Sheaf consistency structure pseudo-section assignment}
\label{tab:globalSectionAssignments}
\begin{tabular}[t]{lcccc}
\hline \hline
            & C1 & C2 & S1 & S2 \\
\hline
Assignments  &  [200,30] & [200,30] & 12.91 & 12.91 \\  
\hline
\end{tabular}
\end{table}

Then, we give assignments to the vertices, such that all of them can be inferred to match perfectly with each other. The assignments are demonstrated in Table \ref{tab:globalSectionAssignments}, where we have two unique assignments such as $12.91$ and $[200,30]$, in which $200$ stands for the number of motorbikes and $30$ stands for the number of cars and they then can be inferred as exactly $12.91$ by the numbers of emission factors shown in Table \ref{tab:EF}. Because the values are perfectly matched right on the vertices, the values propagated to the higher faces are also consistent and therefore completely matched. This situation is called a global section of the sheaf model or the pseudo-section of the consistency structure of the sheaf. All the consistency thresholds in the consistency filtration equal to $0$ as a result. Hence, the filtration in this case is a dictionary, as such
\begin{lstlisting}[language=python]
    {'C': 0.0,
     'S': 0.0,
     'CS1': 0.0,
     'CS2': 0.0,
     'CS3': 0.0,
     'CS4': 0.0,
     'L': 0.0,
     'I': 0.0,
     'K': 1.8553442084620055e-15
    }
\end{lstlisting}

\subsection{Example 2: Sheaf Data Aggregation with simulated signals}
In this example, there are four nodes that are being used to measure a simulated signal. These nodes consist of two cameras and two sensors. The cameras have a sampling frequency rate of once every 600 seconds, while the sensors have a sampling frequency rate of once every 15 seconds. The purpose of this example is to examine how using a sheaf can provide better data fusion (combining multiple data sources to obtain a more accurate result) than simply averaging the data from the different nodes (a method known as "naive data aggregation").

\begin{figure}[h!]
  \centering
  \includegraphics[width=0.65\linewidth]{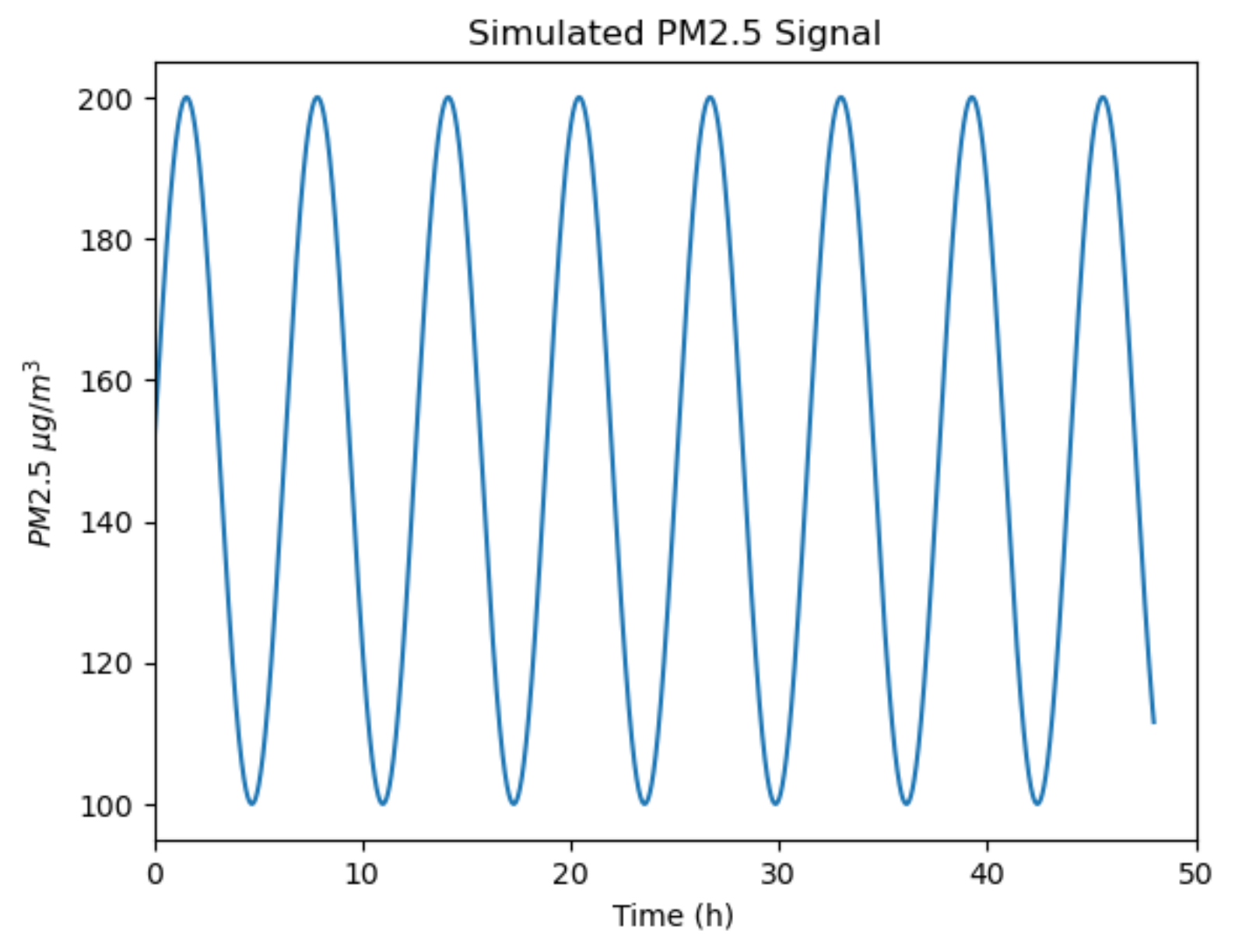}
  \caption{Simulated PM2.5 signal by using a sinusoidal function that varies over time.}
  \label{fig:simulatedPM}
\end{figure}

\begin{figure}[h!]
  \centering
  \includegraphics[width=0.8\linewidth]{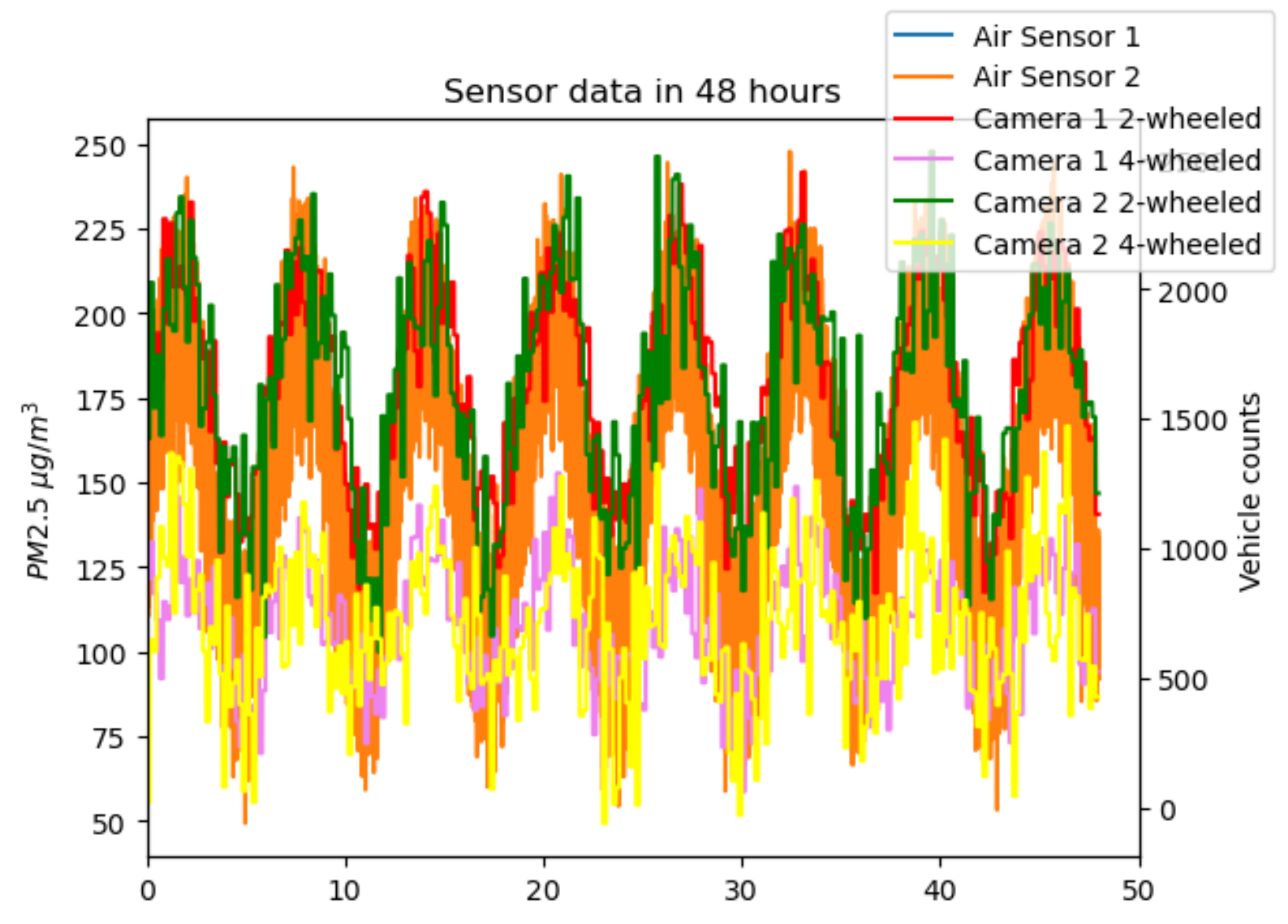}
  \caption{Simulated PM2.5 sensor measurement over time.}
  \label{fig:sensorsPM}
\end{figure}

We create a simulation of the actual PM2.5 signal by using a sinusoidal function that varies over time, with the unit of time being seconds, and limited in the range between $100$ and $200$ by incorporating coefficients to the sinusoidal function, such as $y = 50sin(x) + 150$. The simulation covers a period of 48 hours, illustrated in Fig. \ref{fig:simulatedPM}. Then, to simulate the measurement of the sensor signals, we add different levels of Gaussian noise to each type of sensor. The noise levels for the sensor nodes C1, C2, S1, and S2 are $2.8\%$, $8.3\%$, $11.7\%$, and $16.9\%$, respectively. Additionally, the sensors have different resolutions, so their measurements are not in sync. To ensure that we can propagate the values across the sheaf faces, we need to fill in the values for the vertices every time we update the measurements. To handle the misalignment of the measurements, we assign the previous values to the vertices that have not received a new measurement when we update. The data from the sensors are depicted in Fig. \ref{fig:sensorsPM}.

\begin{figure}[h!]
  \centering
  \includegraphics[width=0.75\linewidth]{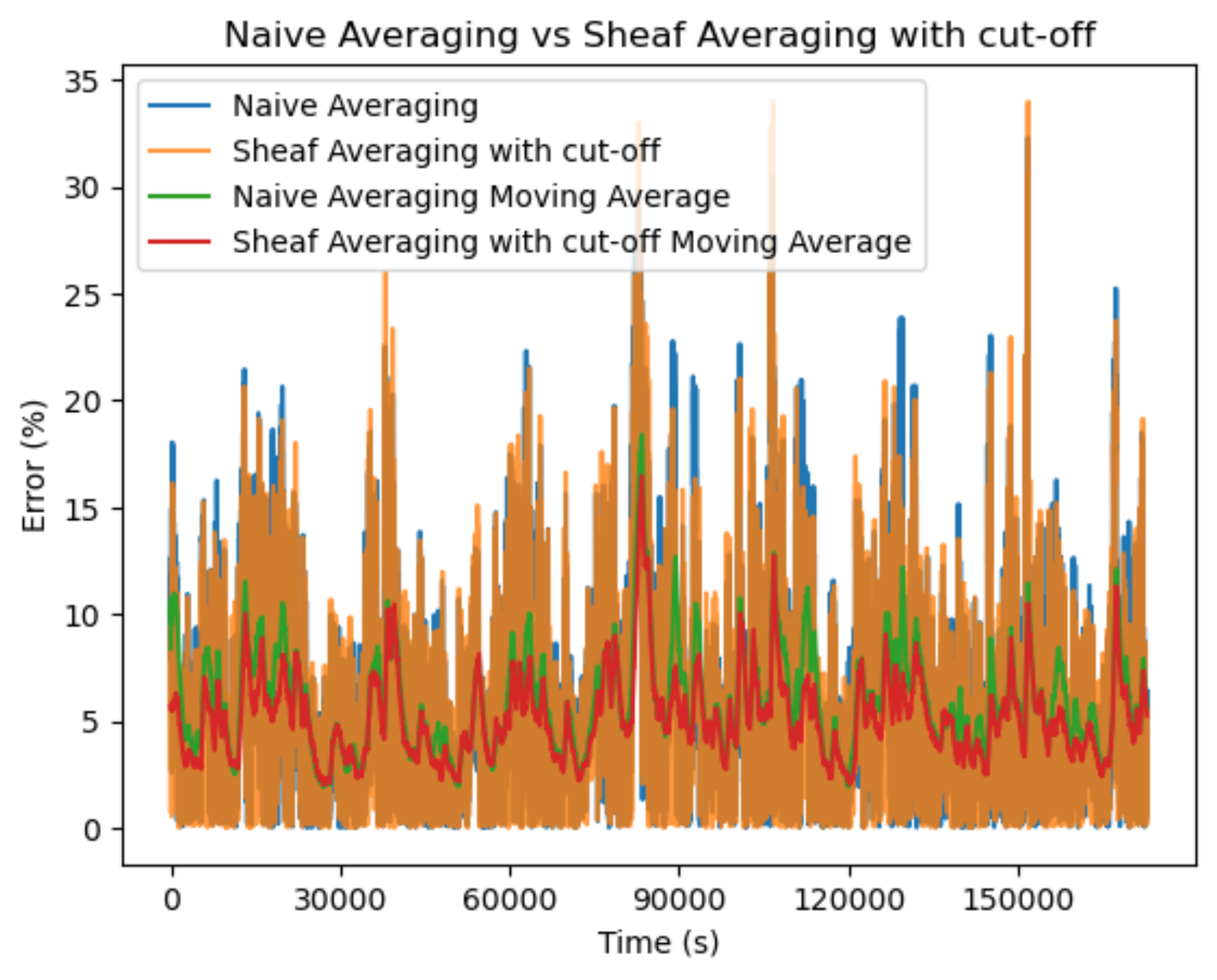}
  \caption{Naive Averaging v.s. Sheaf Averaging with cut-off.}
  \label{fig:sheafVSNaive}
\end{figure}

Given all the aforementioned simulated sensor signals, we conduct a comparison between a naive data averaging and data fusion using sheaf data averaging with cut-off. Specifically, in the naive approach, every time a sensor gets updated, the values of all four sensors are averaged out to have a common value among the sensors at that moment. Unlikely, in our sheaf modeling system, only the consistent faces, which are defined by applying a certain threshold (here, it is the mean value plus half a standard deviation) to the consistency filtration, are considered in the data aggregation process. Both approaches are shown to improve the overall errors, determined by the Mean Absolute Percentage Error (MAPE) metrics shown in Eq. \ref{eq:MAPE}, compared to the individual sensors' error in Table \ref{tab:errorcomparison}, but sheaf with cut-off has further decreased the error of the naive averaging approach by $12.43\%$. To present the result more clearly, Fig. \ref{fig:sheafVSNaive} illustrates the errors of both approaches over time with their moving average. The figure again depicts that values provided by the sheaf (the orange line) are lower than the naive averaging approach (the blue line) over a long course of the simulation. This trend is more visible in their moving average counterpart, where the green line standing for the moving average of the naive averaging values always lie above the red line standing for that of the sheaf average with cut-off approach.

\begin{equation}
\label{eq:MAPE}
MAPE = \frac{1}{N}\sum_{i=1}^N \frac{|y - \hat{y}|}{y} \times 100 \qquad 
\end{equation}

\begin{table}\centering
\caption{Average errors of individual sensors and two data aggregation approaches}
\label{tab:errorcomparison}
\begin{tabular}[t]{lcccccc}
\hline \hline
            & C1 & C2 & S1 & S2 & Naive Averaging & Sheaf Averaging with cut-off \\
\hline
MAPE (\%)  &  11.74\% & 16.93\% & 2.83\% & 8.25\% & 5.92\% & 5.18\% \\  
\hline
\end{tabular}
\end{table}

\begin{figure}[h!]
  \centering
  \includegraphics[width=0.75\linewidth]{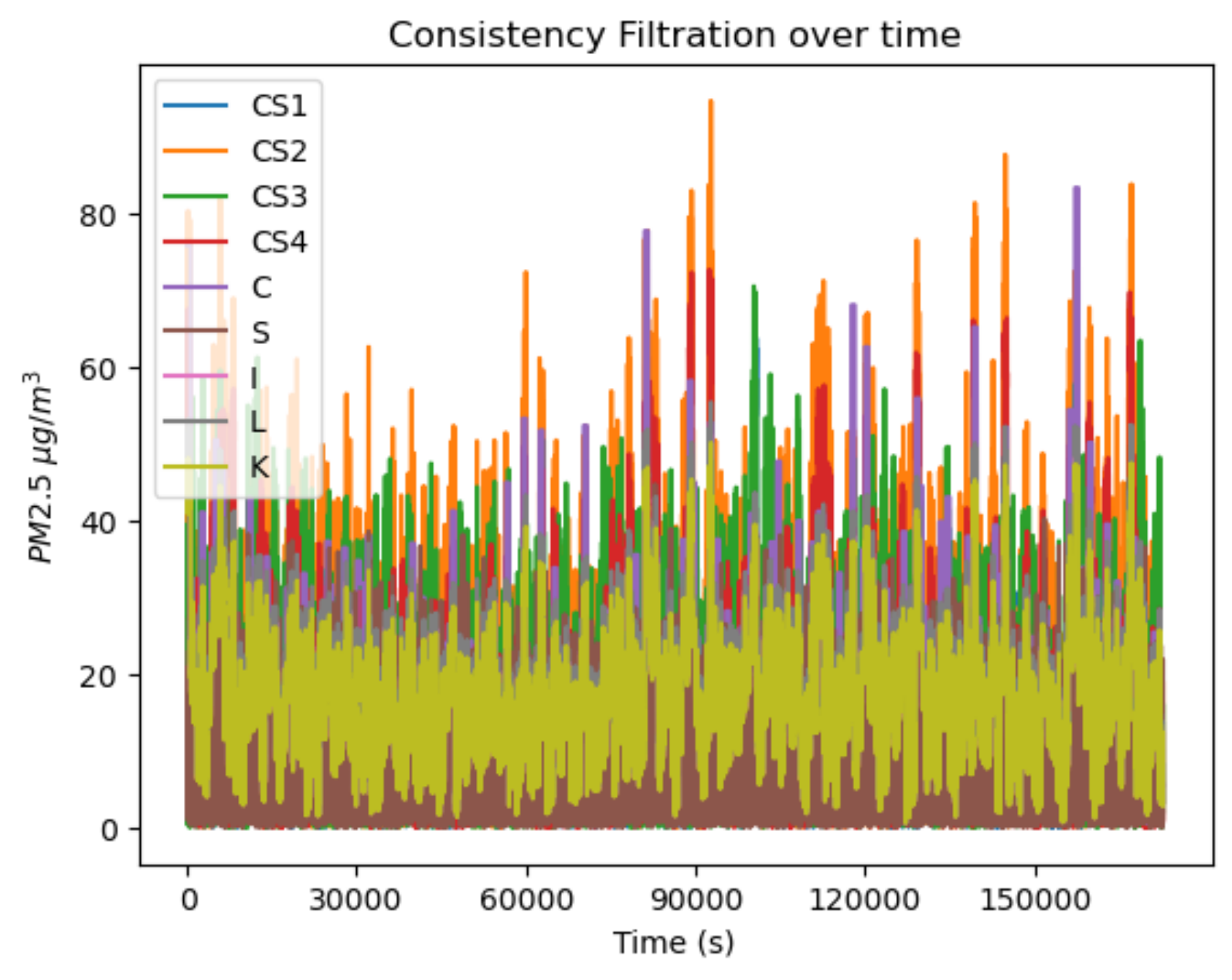}
  \caption{Sheaf filtration of the sensor network over time.}
  \label{fig:sheafFiltration}
\end{figure}

Apparently, by cutting off the inconsistent face out of the consistency structure of the sheaf model would reduce the overall errors of the system. We would like to demonstrate how cutting off inconsistent faces would decrease the error by analyzing the most spread filtration among the consistency filtration sets of the sheaf model over the simulation time, illustration in Fig. \ref{fig:sheafFiltration}. 

\subsection{Example 3: Sheaf Filtration at the minute 1545.5}
\begin{figure}[h!]
  \centering
  \includegraphics[width=0.75\linewidth]{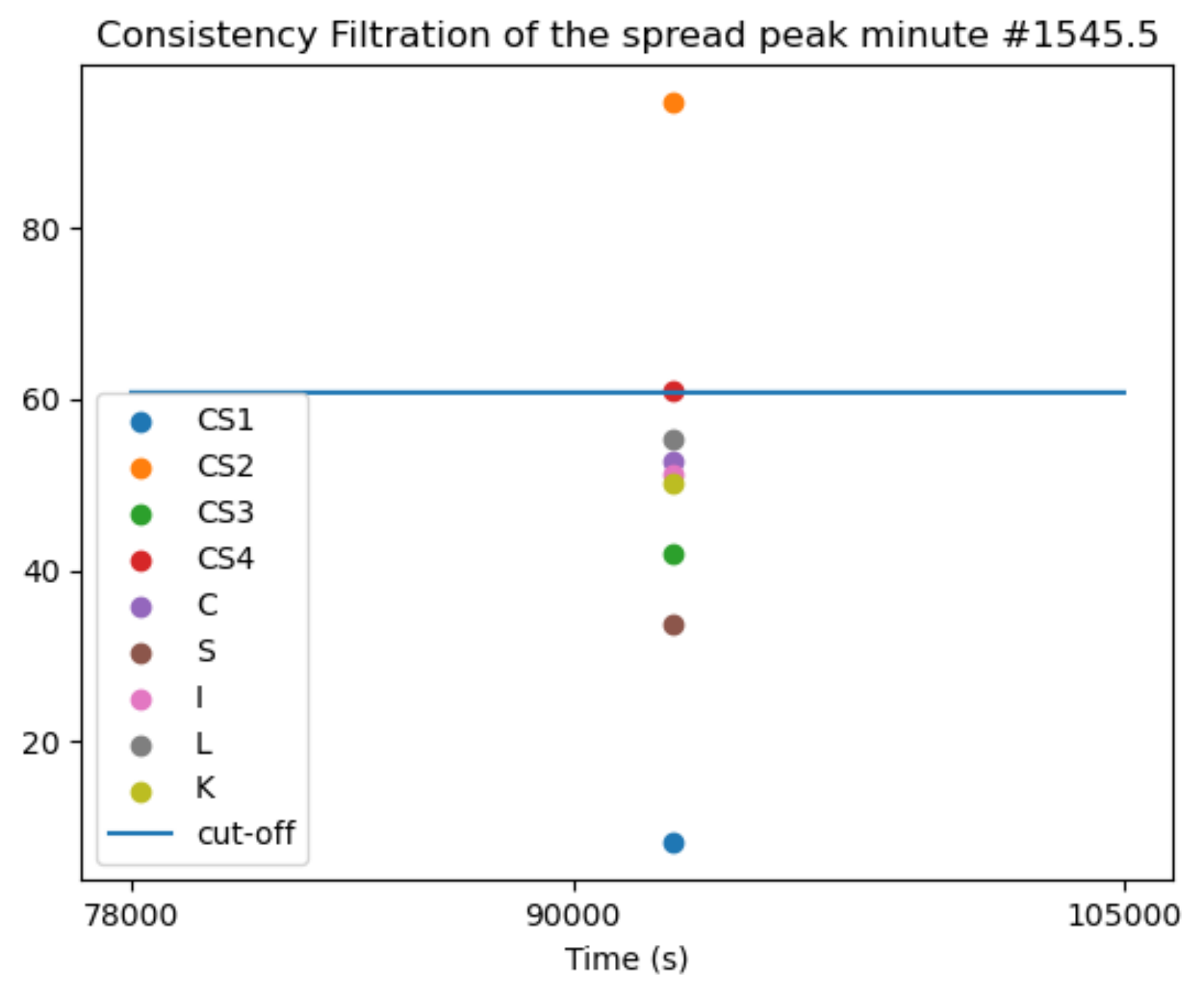}
  \caption{Sheaf filtration spread of the sensor network at the minute 1545.5.}
  \label{fig:sheafFiltrationSpread}
\end{figure}

In Fig. \ref{fig:sheafFiltration}, over time, the consistency of the filtration tends to align with the errors that each sensor produces. Usually, the face sensors $CS2$, $CS4$, and $C$ are the most prone to errors, and these correspond to the two least accurate nodes, $C1$ and $C2$. At the second $\#92730$, or the minute $\#1545.5$, there is a maximum in the data spread as measured by the consistency filtration. We would like to demonstrate the impact of removing faces with high consistency thresholds. At that point of time, the filtration is a dictionary, as such
\begin{lstlisting}[language=python]
    {'CS1': 8.283882338353894,
    'S': 33.620082580239966,
    'CS3': 41.903964918593864,
    'K': 50.105373654118516,
    'I': 51.284484945400095,
    'C': 52.72299262385046,
    'L': 55.393574797089926,
    'CS4': 61.00687496220436,
    'CS2': 94.62695754244432}
\end{lstlisting}
, where the cut-off point is defined as the mean plus half a standard deviation of the filtration distribution, which is in this case equalled to $60.71$. Figure \ref{fig:sheafFiltrationSpread} illustrates the faces in the filtration along with their values and also the cut-off value in order to visualize how the distribution is spread.

Visually, it seems that only the face $CS2$ is eliminated off the sheaf averaging process, but the threshold actually is $60.71$, which means that the face $CS4$ lying exactly on the threshold line but it has the value of $61.0$ and thus is also eliminated. This gives the results in Table \ref{tab:filtrationerrorcomparison}, where sheaf averaging without the cut-off is exactly the same with the naive averaging approach. This again stresses the importance of using the consistency filtration to filter out the inconsistency faces off the averaging process.

\begin{table}\centering
\caption{Filtration errors assessment}
\label{tab:filtrationerrorcomparison}
\begin{tabular}[t]{lccc}
\hline \hline
            & Naive Averaging & Sheaf Averaging without cut-off & Sheaf Averaging with cut-off\\
\hline
MAPE (\%)  &  11.03\% & 11.03\% & \textbf{8.05\%}  \\  
\hline
\end{tabular}
\end{table}

\section*{Acknowledgement}
We would like to express our sincere gratitude to Minh Cong Dinh and Nhien Hao Truong for their assistance in preprocessing video data for vehicle counting. Their valuable contributions were instrumental in the success of our project. We would also like to thank Dr. Victor Rodrigo Iza-Teran for his inspiration and guidance in the use of sheaf-based numerical analysis. His insights and expertise were invaluable in our research. We are deeply grateful to all of these individuals for their support and dedication to our work.

\bibliographystyle{unsrtnat}
\bibliography{references}  





\appendixtitleon
\appendixtitletocon
\begin{appendices}
  \section{Basic Sheaf Python Implementation}
  In this section, we will discuss the various Python code snippets that we use to build the sheaf consistency structure for our systems. These snippets are used in both the Toy Examples section and in our actual, real-world system. Complete snippets could be found at our github repository: https://github.com/a11to1n3/AirSheaf along with the jupyter notebook that shows how to reproduce the toy examples in Section \ref{sec:toyEx}.
  \subsection{Vertex}
      \begin{lstlisting}[language=Python]
from typing import Tuple, List, Dict, Callable, Union
import numpy as np
import networkx as nx

class Vertex(FaceBase):
    def __init__(self,
                name: str,
                value: float = 0,
                ) -> None:

        super().__init__(name, value, 0)
        self._is_connected_to = []
        self._is_subface_of = []
        self._restriction_to_superface = {}

    def connect_to_vertex(self, vertex) -> None:
        self._is_connected_to.append(vertex)
    
    def get_connected_vertices(self) -> List:
        return self._is_connected_to
    
    def get_subfaces(self) -> List:
        return []
      \end{lstlisting}
  \subsection{Face}
  \begin{lstlisting}[language=Python]
class Face(FaceBase):
    def __init__(self,
                name: str,
                value: List = [],
                d: int = 1,
                ) -> None:

        super().__init__(name, value, d)
        self._is_connected_to = []
        self._is_subface_of = []
        self._is_superface_of = []
        self._restriction_to_superface = {}
        self._consistency_threshold = 0
    
    def connect_to_face(self, face) -> None:
        self._is_connected_to.append(face)

    def get_connected_faces(self) -> List:
        return self._is_connected_to

    def set_subface(self, subface) -> None:
        self._is_superface_of.append(subface.get_name())

    def get_subfaces(self) -> List:
        return self._is_superface_of

    def set_consistency_threshold(self, thres) -> None:
        self._consistency_threshold = thres

    def get_consistency_threshold(self) -> float:
        return self._consistency_threshold
      \end{lstlisting}
  \subsection{Sheaf}
  \label{code:sheaf}
  \begin{lstlisting}[language=Python]
class Sheaf:
    def __init__(self) -> None:
        self.faces = {}
        self.consistency_filtration = {}
        self.values = []

    def _set_vertex(self, vertex: type[Vertex]) -> None:
        if "0-face" in list(self.faces.keys()):
            self.faces["0-face"][vertex.name] = vertex
        else:
            self.faces["0-face"] = {}
            self.faces["0-face"][vertex.name] = vertex
    
    def set_vertices(self, list_of_vertices:List[type[Vertex]]) -> None:
        [self._set_vertex(vertex) for vertex in list_of_vertices]

    def get_vertex(self, vertex_name: str):
        if vertex_name in self.faces["0-face"]:
            return self.faces["0-face"][vertex_name]
        
        raise ValueError

    def _set_face(self, face) -> None:
        if f"{face.d}-face" in list(self.faces.keys()):
            self.faces[f"{face.d}-face"][face.name] = face
        else:
            self.faces[f"{face.d}-face"] = {}
            self.faces[f"{face.d}-face"][face.name] = face

    def set_faces(self, list_of_faces:List[type[Face]]) -> None:
        [self._set_face(face) for face in list_of_faces]

    def get_face(self, face_name: str):
        for i in range(1, len(self.faces)):
            if face_name in self.faces[f"{i}-face"]:
                return self.faces[f"{i}-face"][face_name]

        raise ValueError

    def _check_faces_exist(self, faces: List[Union[type[Face],type[Vertex]]]) -> bool:
        if f"{faces[0].d}-face" not in list(self.faces.keys()) and not set(faces).issubset(list(self.faces.values())):
            return False

        return True

    def set_higher_faces(self,
                          list_of_superfaces: Dict[str, Union[Dict[str, Dict[str, Union[type[Face],type[Vertex]]]],Dict[str, Dict[str, Callable]]]]
                        ) -> None:

        for superface in list_of_superfaces:
            assert self._check_faces_exist(list(list_of_superfaces[superface]["subfaces"].values())), "One/All of the faces are not existed in Sheaf"
            combined_higher_face = Face(superface, d=list(list_of_superfaces[superface]["subfaces"].values())[0].d + 1)
            for subface in list_of_superfaces[superface]["subfaces"]:
                combined_higher_face.set_subface(list_of_superfaces[superface]["subfaces"][subface])
                list_of_superfaces[superface]["subfaces"][subface].set_superface(combined_higher_face, list_of_superfaces[superface]["restriction_map"][subface])
            self._set_face(combined_higher_face)

    def _update_vertices(self, value_of_vertices:Dict[str, Union[Tuple[float,float], Tuple[float,Tuple[float]]]]) -> None:
        for vertex in value_of_vertices:
            self.faces["0-face"][vertex].set_value(value_of_vertices[vertex][1:])

    @staticmethod
    def _calculate_value_spread(data_list: List[float]):
        if type(data_list[0]) == list and len(data_list[0]) == 1:
            return np.sqrt((np.cov([data[0] for data in data_list])))
        elif type(data_list[0]) == list:
            return np.sqrt(np.trace(np.cov(np.array(data_list).T)))

        return np.sqrt(np.sum(np.cov(data_list)))

    def _update_consistency_filtration(self, face, consistency_threshold):
        self.consistency_filtration[face] = consistency_threshold

    def _reset_consistency_filtration(self):
        self.consistency_filtration = {}

    def get_consistency_filtration(self) -> Dict[str, float]:
        return self.consistency_filtration

    def _propagate_level_by_level(self) -> None:
        self._reset_consistency_filtration()
        for d in self.faces:
            if d != "0-face":
                for face in self.faces[d]:
                    value_list = []
                    running_faces = [self.faces[d][face]]
                    subfaces = {}
                    subfaces[face] = []
                    paths = []
                    for i in range(self.faces[d][face].get_face_level()-1,-1,-1):
                        if len(running_faces) == 0:
                            break
                        
                        temp = []
                        while len(running_faces) != 0:
                            curr_face = running_faces.pop(0)
                            for f in self.faces[f"{i}-face"]:
                                if i > 0:
                                    fa = self.get_face(f)
                                else:
                                    fa = self.get_vertex(f)
                                if fa.get_face_level() != 0 and curr_face.name in fa.get_direct_superfaces():
                                    subfaces[curr_face.name].append(fa.name)
                                    temp.append(fa)
                                    subfaces[fa.name] = []
                                elif fa.get_face_level() == 0 and curr_face.name in fa.get_direct_superfaces():
                                    subfaces[curr_face.name].append(fa.name)
                                    back = [curr_face.name]
                                    path = fa.name+f"-{curr_face.name}"
                                    while len(back) != 0:
                                        curr = back.pop(0)
                                        for key, val in subfaces.items():
                                            if curr in val:
                                                back.append(key)
                                                path += f"-{key}"
                                    paths.append(path)

                        else:
                            running_faces = temp
                    
                    for path in paths:
                        nodes = path.split("-")
                        value = self.get_vertex(nodes[0]).value
                        for idx in range(len(nodes)-1):
                            if idx == 0:
                                subface = self.get_vertex(nodes[idx])
                            else:
                                subface = self.get_face(nodes[idx])

                            value = subface.lift_up_to_superface_func(nodes[idx+1])(value)
                        
                        if type(value) == list and len(value) == 1:
                            value_list.append(value[0])
                        else:
                            value_list.append(value)

                    self.faces[d][face].set_value(value_list)
                    consistency_threshold = self._calculate_value_spread(value_list)
                    self.faces[d][face].set_consistency_threshold(consistency_threshold)
                    self._update_consistency_filtration(face, consistency_threshold)

    def propagate(self, value_of_vertices:Dict[str, Union[float, Tuple[float]]]) -> None:
        self._update_vertices(value_of_vertices)
        self._propagate_level_by_level()
        self.consistency_filtration = {k: v for k, v in sorted(self.consistency_filtration.items(), key=lambda item: item[1])}
        self.consistency_radius = self.consistency_filtration[list(self.consistency_filtration.keys())[-1]]
        value_list = []
        for i in range(len(self.consistency_filtration)):
            if list(self.consistency_filtration.values())[i] <= np.mean(list(self.consistency_filtration.values())) + 0.5*np.std(list(self.consistency_filtration.values())):
                face = self.get_face(list(self.consistency_filtration.keys())[i])
                value_list.append(np.mean(face.value))

        self.value = np.mean(value_list)

    def visualize_sheaf_as_digraph(self) -> type[nx.MultiDiGraph()]:
        G = nx.MultiDiGraph()
        for d in range(len(self.faces)):
            for face in self.faces[f"{d}-face"]:
                G.add_node(face)
                subfaces = self.faces[f"{d}-face"][face].get_subfaces()
                if len(subfaces) > 0:
                    for subface in subfaces:
                        G.add_edge(subface, face)

        return G
      \end{lstlisting}
\end{appendices}

\end{document}